\documentclass[tradiabstract]{aa} 
\usepackage{graphicx}
\usepackage{txfonts}
\def\HeI{He\,{\sc i}}
\def\HeII{He\,{\sc ii}}

\def\NII{N\,{\sc ii}}
\def\NIII{N\,{\sc iii}}

\def\SiIII{Si\,{\sc iii}}

\def\Lstar{$L_{\ast}$}
\def\kms {km~s$^{-1}$}

\def\Teff{$T_{\rm eff}$}

\def\Rsun {$R_{\odot}$}

\def\Mdot{${\dot M}$}

\def\vinf {$v_{\rm \infty}$}

\def\fv{$f_{\rm v}$}

\begin{document}
\title{On the nature of candidate luminous blue variables in M33\thanks{
Based on observations made with the William Herschel Telescope
operated on the island of La Palma by the Isaac Newton Group 
in the Spanish Observatorio del Roque de los Muchachos 
of the Instituto de Astrof\'{\i}sica de Canarias.}
}
\author{J.~S.~Clark\inst{1}
\and N.~Castro\inst{2}
\and M.~Garcia\inst{3}
\and A.~Herrero\inst{3}
\and F.~Najarro\inst{4}
\and I.~Negueruela\inst{5}
\and B.~W.~Ritchie\inst{1}
\and K.~T.~Smith\inst{6}}
\institute{
$^1$Department of Physics and Astronomy, The Open 
University, Walton Hall, Milton Keynes, MK7 6AA, UK\\
$^2$Institute of Astronomy \& Astrophysics, National Observatory of Athens, I.
Metaxa \& Vas. Pavlou St., P. Penteli, 15236 Athens, Greece\\
$^3$Departamento de Astrofisica, Universidad de La Laguna, 38205, La Laguna 
Tenerife, Spain\\
$^4$Departamento de Astrof\'{\i}sica, Centro de Astrobiolog\'{\i}a, 
(CSIC-INTA), Ctra. Torrej\'on a Ajalvir, km 4,  28850 Torrej\'on de Ardoz, 
Madrid, Spain\\
$^5$Departamento de F\'{i}sica, Ingenar\'{i}a de Sistemas y Teor\'{i}a de la Se\~{n}al, Universidad de Alicante, Apdo. 99,
E03080 Alicante, Spain
$^6$School of Chemistry, The University of Nottingham, University park, Nottingham, NG7 2RD, UK 
}

   \abstract{Luminous blue variables (LBVs) are a class of highly unstable stars that have been proposed to play a critical role in massive stellar       
evolution as well as being the progenitors of some of the most luminous supernovae known. However the physical processes underlying 
 their characteristic instabilities are currently unknown.
            In order to provide observational constraints on this behaviour we have initiated a pilot study of the population of 
(candidate) LBVs in the Local Group galaxy M33.       
            To accomplish this we have obtained new spectra of 18 examples within M33. These provide a baseline of $\geq$4~yr with respect to previous 
observations, which is well suited to identifying LBV outbursts. We also employed existing multi-epoch optical and mid-IR surveys of M33 to further 
constrain 
the variability of the sample and search for the presence of dusty ejecta.
            Combining the datasets reveals that spectroscopic and photometric variability appears common, although 
 in the majority of cases further observations will be needed to distinguish between an origin for this behavour in 
short lived stochastic wind structure and low level photospheric pulsations or coherent long term LBV excursions. Of the
 known LBVs we report a  hitherto unidentified  excursion of M33 Var C 
between 2001-5, while the transition of the WNLh 
star B517 to a cooler B supergiant phase between 1993-2010 implies an LBV classification. Proof-of-concept quantitative 
model atmosphere analysis is provided  for Romano's star; the  resultant stellar parameters being consistent with the finding 
  that the LBV excursions of this star are accompanied by changes in bolometric luminosity.  The combination of temperature and luminosity of two
stars, the BHG [HS80] 110A and the cool hypergiant B324, appear to be in violation of the empirical Humphreys-Davidson limit.
Mid-IR observations demonstrate that a number of candidates appear associated with hot circumstellar dust, although no objects as extreme as $\eta$ Car are 
identified. The 
combined dataset suggests that the criteria employed to identify candidate LBVs results in a heterogeneous sample, also containing stars demonstrating the B[e] 
phenomenon. Of these, a subset of
 optically faint, low luminosity stars associated with  hot dust are of particular interest since they appear similar to the likely progenitor of SN 2008S and the 2008 
NGC300 transient (albeit suffering less intrinsic extinction).
           The results of such a multiwavelength observational approach, employing multiplexing spectrographs and supplemented with quantitative model atmosphere analysis,
appears to show considerable promise in  both identifying and characterising the physical properties of LBVs as well as other short lived phases of massive 
stellar evolution.}


\maketitle

\section{Introduction}

The advent of recent large  transient surveys and deep, multiwavelength 
observations of the stellar content of external galaxies raises the possibility of opening up  the final stages of massive 
stellar evolution to unbiased, quantitative study. Since such objects both drive the evolution of their host galaxies while 
serving as the progenitors of a number of energetic phenomena, including core collapse supernovae (SNe), gamma ray bursts  and 
X- and  $\gamma$-ray binaries, an understanding of their brief and violent lifecycle is of wide interest. 
While it had been supposed that mass loss via (radiatively driven) stellar winds drove the evolution from the main sequence 
through the H-depleted Wolf Rayet (WR), recent observational and theoretical developments suggest that this in fact might
be an oversimplification and that other agents - specifically binary interaction and impulsive mass loss events - 
 may also contribute to this process. 

Regarding the former, extensive radial velocity surveys have revealed both a high binary
 fraction amongst massive stars (e.g. Sana \& Evans \cite{sana}, Ritchie et al. in prep.) and evidence for distinct binary evolutionary channels (e.g.
Clark et al. \cite{clark11a}) confirming the predictions of previous theoretical studies (e.g. Petrovic et al. 
\cite{petrovic}). The latter suggestion implicates the mass loss episodes associated with the 
transitional  luminous blue variables (LBVs) in  mediating the transition to WRs (Smith \& Owocki \cite{smith06}). Observations of Galactic LBVs demonstrate
 that a subset are associated with massive dusty circumstellar nebulae ($\sim$0.01-15~$M_{\odot}$;
 Clark et al. \cite{clark03}) which have historically been thought to arise in the so-called `giant' eruptions; events characterised by
 increases in  bolometric luminosity ($L_{\rm bol}$) and which help distinguish LBVs from other luminous evolved stars.
 (e.g. Humphreys \& Davidson \cite{hd}). However, recent observations demonstrate that 
such outbursts appear to be strikingly diverse and are not always accompanied by significant mass loss (e.g. Clark et al. 
\cite{clark09}).

Given the possible role played by LBVs in the evolution of massive stars, there is considerable interest in observationally 
constraining the nature of the outbursts and hence the underlying physics driving these instabilities. Unfortunately, the LBV 
phase appears to be brief, resulting  in only a handful of widely recognised Galactic examples (Clark et al. \cite{clark05a}).
However,  recent IR galactic plane surveys have greatly expanded on the number of known candidates 
(Clark et al. \cite{clark03}, Gvaramadze et al. \cite{gvaramadze}, Wachter et al. \cite{wachter}, Stringfellow et al. 
\cite{stringfellow}). Such objects are 
invaluable since their spatially resolved circumstellar ejecta comprise a record of their mass loss history, although 
 their distribution throughout the plane means that long term monitoring to determine their duty cycles is an observationally expensive
proposition.

An alternative approach is to observe candidates in external galaxies. While such a strategy complicates the spatial separation of stellar and 
nebular emission, it is observationally efficient given current 4-8~m class telescopes and associated 
multiplexing spectrographs. Moreover, such an approach, informed by ongoing SNe monitoring campaigns has the potential to 
efficiently identify new examples of giant LBV eruptions for detailed follow ups; the so-called `SNe imposters'. In order to 
 complement  our ongoing program of observations of Galactic LBVs, we have recently undertaken a pilot study of 
(candidate) LBVs in M33 to assess the efficacy of this strategy.  

To accomplish this we obtained new 4000-7000~{\AA} spectroscopic observations of 17 (candidate) LBVs (Table 1) selected from 
the seminal paper by Massey et al. (\cite{massey07}; henceforth 
Ma07); the resultant $\geq$4~yr baseline between observations well matching the characterstic timescale of LBV excursions. 
The S/N,  spectral resolution and wavelength coverage of the resultant spectra  - encompassing H$\alpha$, a prime mass loss diagnostic, 
as well as temperature and metallicity diagnostics between 4000-5000~{\AA} - in principle permits  non-local thermal equilibrium (non-LTE) 
model atmosphere analysis of the stars in question. We were also able to supplement this  dataset 
with three additional unpublished spectra obtained in 2003 at the  W. M. Keck observatory and which are further described in   Sect. 2. Of 
these spectra two were of targets already included in our WHT target list and one - M33 Var B - was new.

Moreover, we may also make use of the wealth of existing data on the stellar population of  M33
in order to investigate both the photometric history of, and the presence (or otherwise) and properties 
of  dusty circumstellar ejecta associated with  the (candidate) LBVs. 
For this facet of the program we utilised the full census of such stars published by Ma07 (their Table 18, not reproduced for brevity, 
which contains all 18 objects for which 
we obtained spectroscopy), supplemented with the remaining WN9-11h and 
B super-/hyper-giants within M33 as well as additional cLBVs from a variety of sources; the additional stars 
and associated references are listed  in Table 2\footnote{In the absence of a confirmatory spectrum we currently exclude the candidate reported by
Burggraf \& Weis (\cite{burggraf08}).}.

Consequently, the paper is structured as follows. Data acquisition and reduction is described in Sect. 2, with the resultant spectra 
discussed in  Sect. 3, with the full dataset present online in appendix A. Photometric data are presented in Sect. 4 and 
we discuss the nature of the candidate LBVs in Sect. 5, before summarising and placing our 
results in a wider context in Sect. 6.

\begin{figure}[h]
\includegraphics[width=7cm,angle=270]{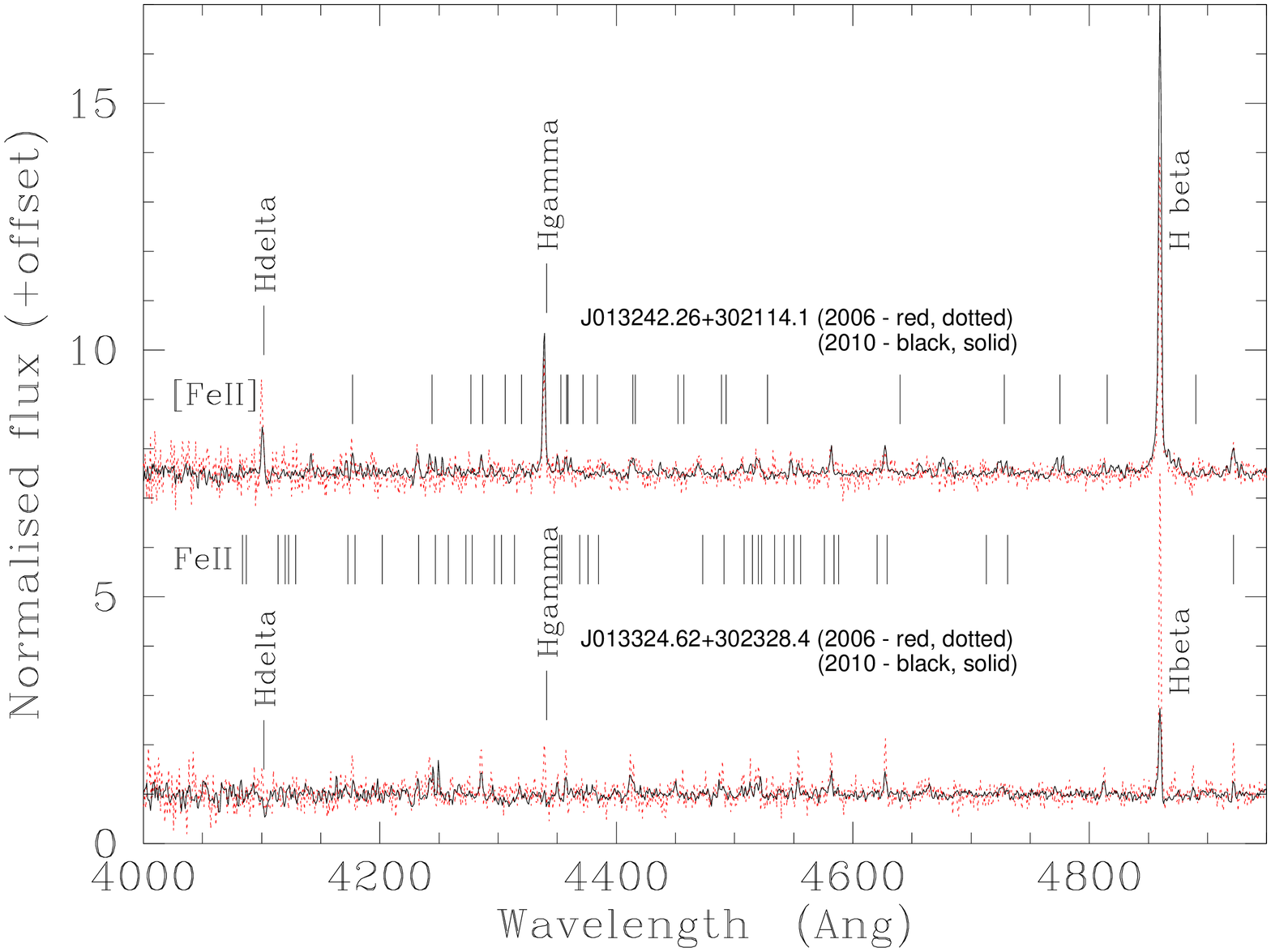}
\includegraphics[width=7cm,angle=270]{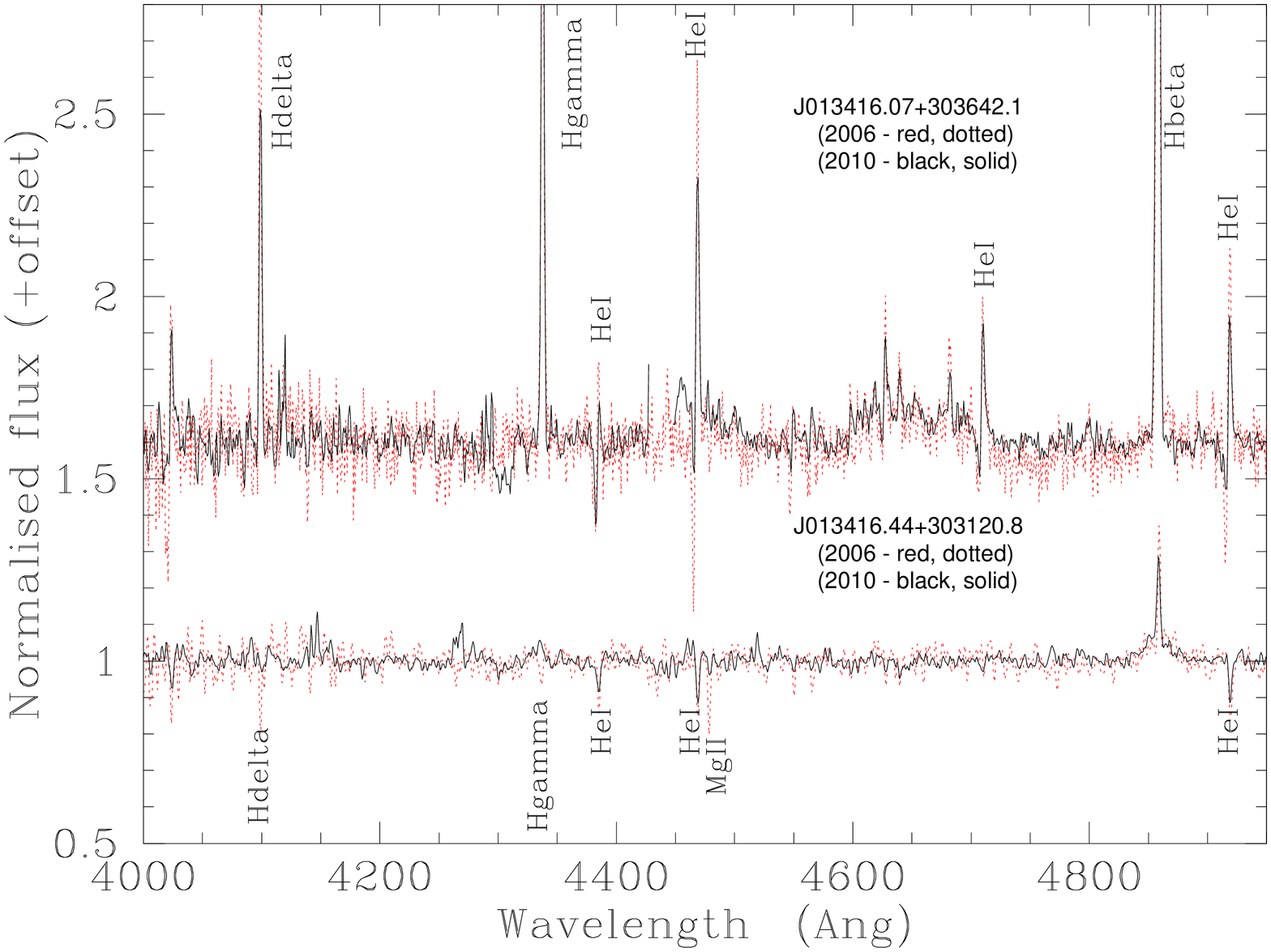}
\caption{Multiepoch optical spectra of selected stars {\em Top 
panel} -  illustrative  Iron stars and {\em bottom panel} - 
the P Cygni LBV J013416.07+3-3642.1 and 
the B hypergiant J013416.44+303120.8}
\end{figure}

\begin{figure}[h]
\includegraphics[width=7cm,angle=270]{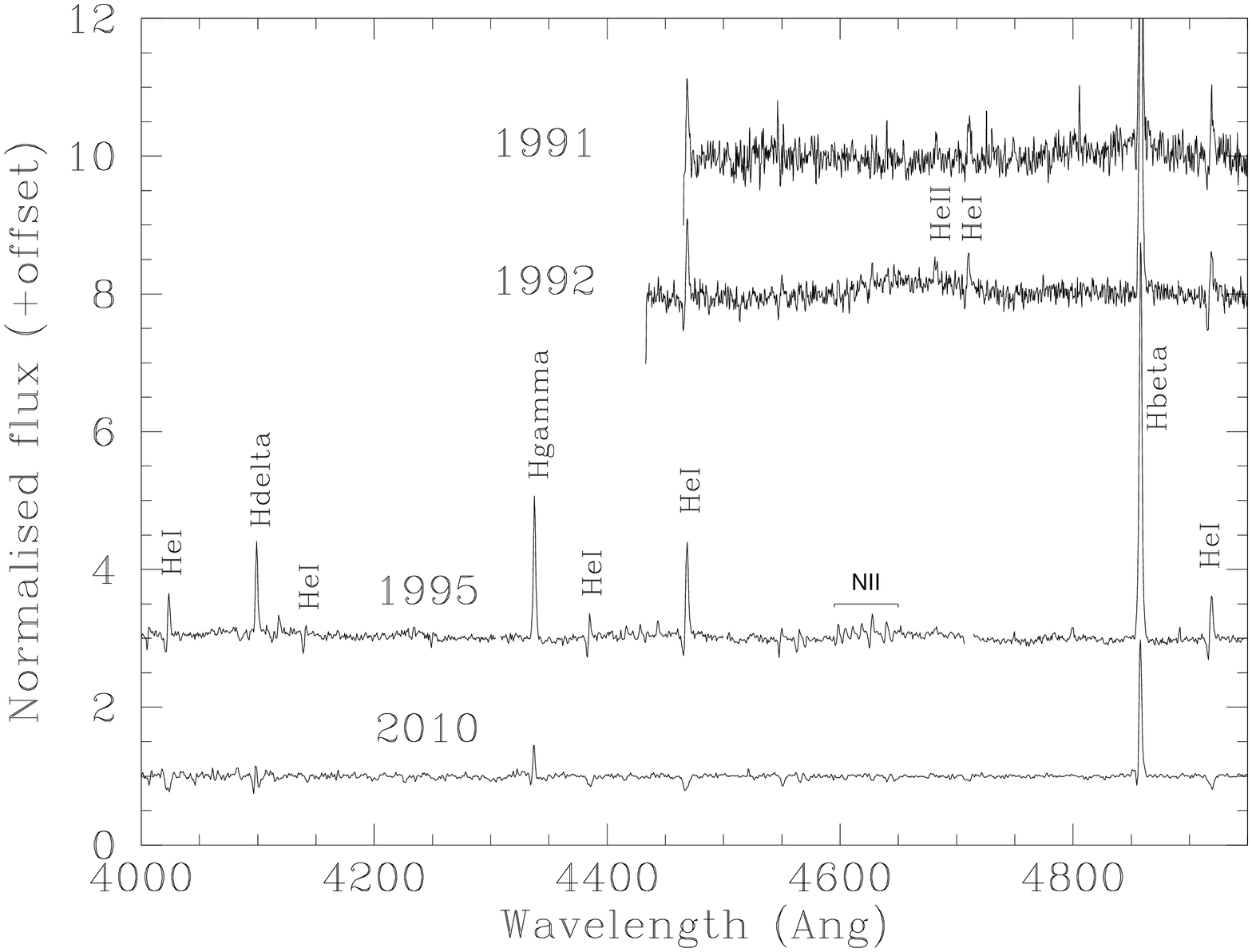}
\includegraphics[width=7cm,angle=270]{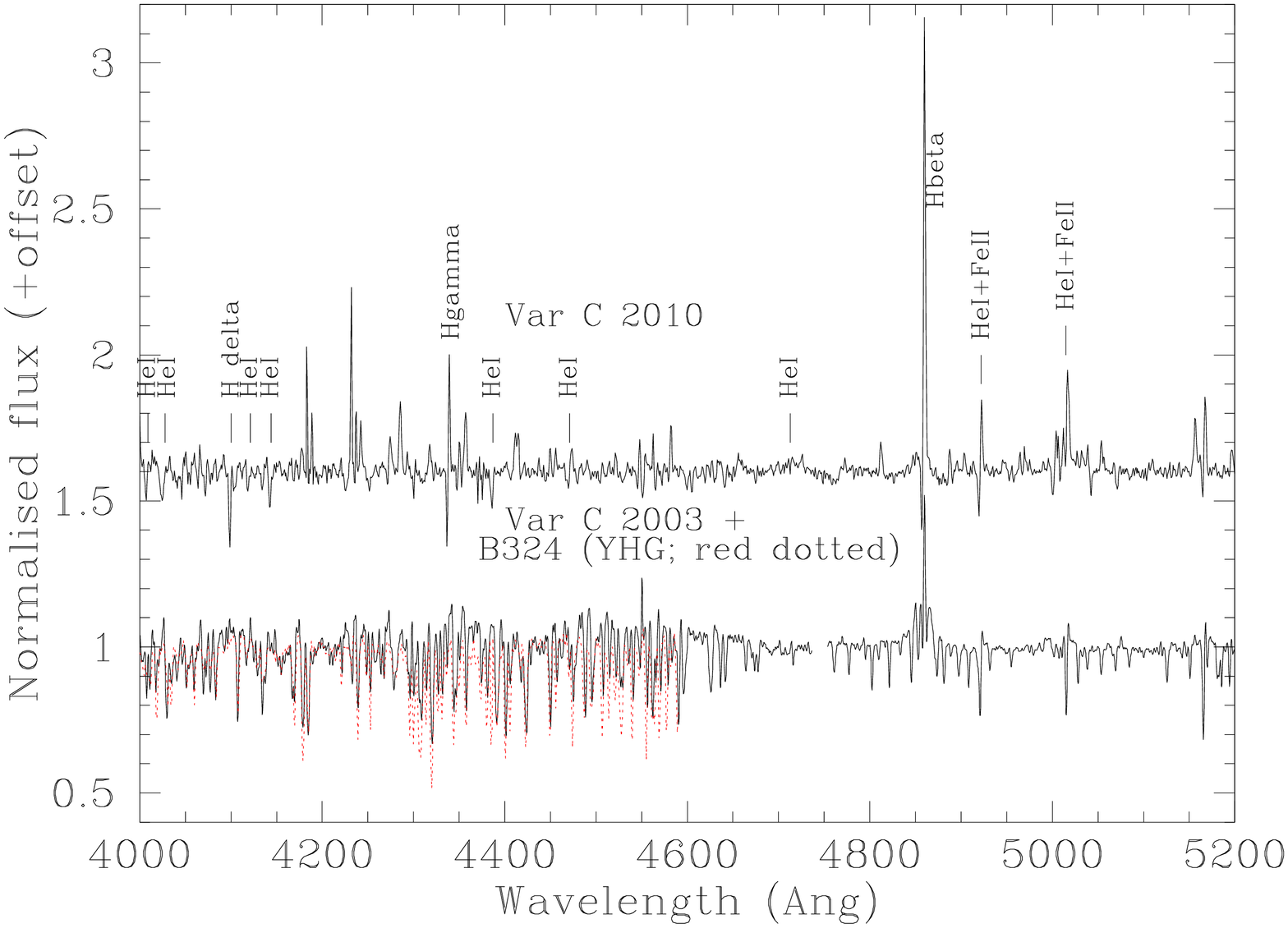}
\caption{Multiepoch optical spectra of selected  LBVs. {\em Top 
panel} - the P Cygni star J013339.52+304540.5; note that the 2010 spectrum is reproduced in Fig. A.3, where weak Si\,{\sc iii}, 
O\,{\sc ii} and C\,{\sc iii} absorption features are visible. {\em Bottom panel}
- the {\em bona fide} LBV M33 Var C, with the spectrum of the cool F0-5Ia$^+$ 
hypergiant B324 overlayed (Monteverde et al. \cite{monteverde} and Sect. 5.3.2).}
\end{figure}

\section{Observations and data reduction}

The spectroscopic observations of 17 of our targets (Table 1) were performed at the 4.2~m William Herschel Telescope (WHT)
in the ``Observatorio del Roque de los Muchachos'' in La Palma, Spain.
We used the AutoFib2 plus Wide Field Fibre Optical Spectrograph (AF2-WYFFOS)
combination for multi-object spectroscopy. 
WYFFOS can be fed with up to 150 1.6$\arcsec$-diameter fibres, 
positioned on the user provided coordinates by the AF2 robot
over a wide ($\sim$1$\deg$ diameter) field of view.

The observing run started on 2010 September 29 and lasted four nights.
We designed two AF2 fibre configurations to cover our targets.
In both cases, we first optimized the number of 
allocated fibers and their priority
(which was assigned based on obtaining a full sample of each sub-category of candidate
LBV defined by Ma07, prior indication of variability and optical magnitude) and 
maximized the number of allocated sky fibers (see below) afterwards.
We used the R1200B and R1200R gratings
to cover the optical-blue spectra ($\sim$ 4000-5000~\AA)
and H${\alpha}$~ region with a resolution of roughly 2~\AA.
The total exposure time for the first configuration
amounted to 9 hours and 6.5 hours with the R1200B and R1200R gratings respectively,
broken into 1800s blocks for cosmic-ray subtraction.
Due to weather conditions, the second configuration
was only observed during  2.5 hours with R1200B;
consequently, the spectra of stars observed in this configuration have much poorer 
signal-to-noise ratios.

The data were reduced following the recipe implemented in the pipeline provided by the AF2 
webpage\footnote{http://www.ing.iac.es/astronomy/instruments/af2/reduction.html} at that time. 
We modified the original procedure to include an initial normalization of the flat-field 
before its correction, and an algorithm for cosmic ray rejection (van Dokkum \cite{vanD}). 
The code then follows the standard steps for multi-fibre spectrum reduction, using 
IRAF\footnote{IRAF is 
distributed by the National Optical Astronomy Observatory, which is operated by the Association of Universities for 
Research in Astronomy, Inc., under cooperative agreement with  the National Science Foundation.} 
tasks: bias, flat-field correction, throughput calibration between fibres, 
wavelength calibration and sky subtraction.  

Due to the fibres head-size and distribution, fibres cannot be placed  closer than 25$\arcsec$ to one another, 
and therefore an accurate and individual sky subtraction is not possible since information
on each star's local background is not available. 
In addition to the fibres positioned on M33 stars, 
a set of supplementary fibres were added to our configurations to obtain 
an average estimation of the sky contribution. 
Subtracting this component allows us to remove both the Earth atmosphere effects 
and any contribution in the M33 field. However, the local nebular contamination 
will still be present in the final reduced stellar spectra, and must be considered in the analysis.
The nebular contamination mainly affects the Balmer lines, 
and usually leaves the rest of relevant lines for morphological classification 
and quantitative analysis unaffected.

Additionally, we were able to make use of unpublished spectra with a resolution of $\sim$1.8{\AA} of M33 Var B and C and 
J013350.12+304126.6 obtained on 2003 November 30 with the
DEIMOS spectrograph at the W. M. Keck observatory; a discussion of data acquisition and 
reduction may be found in Cordiner et al. 
(\cite{cordiner}).

Finally, we also employed publically available optical and mid-IR photometric data 
to study the full M33 candidate LBV census of Ma07, supplemented with the stars in Table 2. These data consist 
of the optical photometric variability surveys of Shporer \& Mazeh (\cite{shporer}) and Hartman et al. (\cite{hartman}) 
as well as the single epoch study of Massey et al. (\cite{massey06}) and the multi-epoch 
mid-IR surveys of  McQuinn et al. (\cite{mcquinn})  and Thompson et al. (\cite{thompson}); these data are analysed in Sect. 4.

\begin{table*}
\begin{center}
\caption[]{Summary of WHT \& Keck spectroscopic target list}
\begin{tabular}{lccccc} 
\hline
\hline
LGGS J\#  & Aliases & Classification & Telescope & $V$  &Last observed \\
\hline
013242.26+302114.1 & - & Iron star & WHT&17.44 &2006 Sept$^a$\\
013324.62+302328.4 & - & Iron star & WHT&19.58 &2006 Sept$^a$\\
013350.12+304126.6 & UIT212 &  Iron star &WHT, Keck &16.82 &1993-5$^b$\\
013406.63+304147.8 & UIT 301 & Iron star & WHT&16.08 &2001$^c$\\
                   & [HS80] B416 &  & &   & \\
\hline
013357.73+301714.2$^*$ & - &  B hypergiant & WHT&17.39 &2006 Sept$^a$\\
013416.44+303120.8$^*$ & - &  B hypergiant & WHT&17.10&2006 Sept$^a$\\
013422.91+304411.0$^*$ & - &  B hypergiant & WHT&17.22 &2006 Sept$^a$\\
013424.78+303306.6$^*$ & - &  B hypergiant & WHT&16.84 &2006 Sept$^a$\\
013429.64+303732.1$^*$ & - &  B hypergiant & WHT&17.10 &2006 Sept$^a$\\
\hline
013339.52+304540.5 & B517 & P Cyg LBVc & WHT&17.50&2006 Sept$^a$ \\
013341.28+302237.2$^*$ & [HS80] 110-A & P Cyg LBVc & WHT&16.28 &2006 Sept$^a$\\
013416.07+303642.1$^*$ & - & P Cyg LBVc & WHT&17.95&2010 Oct$^d$ \\
\hline
013309.14+303954.5 & UIT045 & Ofpe/WNL & WHT&17.91 &1993-5$^b$\\
013327.26+303909.1 & UIT104 & Ofpe/WNL &WHT &17.95 &1993-5$^b$\\
013509.73+304157.3$^*$ & Romano's star &Ofpe/WNL & WHT&18.04 & 2010 Oct$^d$ \\  
\hline
013349.23+303809.1$^*$ & Var B & LBV & Keck &16.21 &2005$^e$ \\                       
013335.14+303600.4$^*$ & Var C & LBV & WHT, Keck&16.43 &2003-5$^e$\\
013416.10+303344.9 & UIT 341 & LBVc & WHT &17.12&1993-5$^{b,f}$\\
                    & [HS80] B526    & &  &         \\
\hline
\end{tabular}
\end{center}
{Summary of  of our WHT and Keck target stars, common aliases, 
classification following the ammended scheme of Ma07 (Sect. 3; noting that Romano's star should 
be now be considered a {\em bona fide} LBV), LGGS V band magnitude (observations made between 2000 November - 2003 February) and  
date of most recent spectroscopic observations in the $\sim$4000-5000{\AA} window
 ( $^{a}$Ma07, 
$^b$Massey et al. \cite{massey96},
$^c$Fabrika et al. \cite{fabrika}, 
$^d$Neugent \& Massey \cite{neugent} 
$^e$Viotti et al. \cite{viotti}, 
$^f$Monteverde et al. \cite{monteverde}). Stars for which
contemporaneous historical observations of the 
$\sim$6000-7000{\AA} window are also available in the literature 
are indicated by an asterix.}\\
\end{table*}

\begin{table}
\begin{center}
\caption[]{Summary of additional candidate LBVs in M33}
\begin{tabular}{lr}
\hline
\hline
Name      &  Reference \\
\hline
M33 Var 2(=VHK2)     & van den Bergh et al. (\cite{83})\\
N93351    &   Valeev et al. (\cite{valeev09}) \\
N45901    &      Valeev et al. (\cite{valeev10}) \\
N125093   &      Valeev et al. (\cite{valeev10}) \\
\hline
UIT005    &    Urbaneja et al. (\cite{urbaneja11}) \\
IFM-B 0301 & Monteverde et al. (\cite{monteverde})\\
IFM-B 0515 & Monteverde et al. (\cite{monteverde})\\
IFM-B 1040 & Monteverde et al. (\cite{monteverde}) \\
IFM-B 1054 & Monteverde et al. (\cite{monteverde}) \\
IFM-B 1330   &  Monteverde et al. (\cite{monteverde})\\
IFM-B 1345 & Monteverde et al. (\cite{monteverde})\\ 
OB 10-10 (=UIT 181) & Massey et al. (\cite{massey95})\\
\hline
LGGS J013354.85+303222.8  & Neugent \& Massey (\cite{neugent}) \\
LGGS J013418.37+303837.0 & Neugent \& Massey (\cite{neugent}) \\
LGGS J013432.24+304702.7 & Neugent \& Massey (\cite{neugent}) \\
LGGS J013432.50+304703.5 & Neugent \& Massey (\cite{neugent}) \\
\hline
\end{tabular}
\end{center}
{{\em Top panel:} supplemental  (candidate) LBVs in addition to those in the census of 
 Ma07 (their Table 18)  noting that Var83 \& B416  are not labeled as such in that work, 
but are LGGS J013410.93+303437.6   and LGGS 
J013406.63+304147.8 respectively. Of these, N45901, N93351 and N125093 may be classified as iron stars (Sect. 3)
while M33 Var 2 has historically been classified as an  LBV. {\em Middle panel:}
BA super- and hypergiants and  {\em bottom panel:} Additional Ofpe/WNL and WN9-11 stars from 
Neugent \& Massey (\cite{neugent}; note that these authors now classify J013416.07+303642.1 as Ofpe/WNL rather than P Cygni-type in Ma07).}\\
\end{table}

\section{Spectroscopy}

Ma07 divide their candidate LBVs into five subsets. Of these, the P Cygni and Ofpe/WNL stars are well defined spectral and evolutionary 
classifications, while the {\em bona fide} LBVs such as M33 Var B and C are also clearly identifiable; we adopt these categories unchanged.
However the remaining groupings - the hot and cool LBV candidates - are more problematic since such nomenclature implies a specific 
evolutionary phase for what are likely heterogeneous sets of objects. 

Motivated by the appearance of M33 Var C in photometric minimum, Ma07 apply a  classification of hot candidate LBV to 
those stars showing strong emission in both the Balmer series and low excitation metallic transitions such as Fe\,{\sc ii}, although they 
recognise that such criteria will also include stars demonstrating the B[e] phenomenon (e.g. Lamers et al. \cite{lamers}). 
This is intriguing, given the still uncertain physical relationship  between both classes of star and the current lack of placement of the 
latter in a coherent evolutionary scheme (see e.g. Zickgraf \cite{zickgraf06} for a review of this topic). Unfortunately,
 the limited spectral resolution and wavelength coverage of the existing  spectra of these stars prevent us from 
inferring the geometry of the circumstellar material via line profiles (e.g. Zickgraf et al. \cite{zickgraf85}) or H\,{\sc i} line ratios
(Lenorzer et al. \cite{lenorzer}) and hence distinguishing between both possibilities. Moreover the Fe\,{\sc ii} emission line spectra of
LBVs are also highly variable (e.g. AG Car, HR Car,  R71 and  R127; Stahl et al. \cite{stahl01}, Szeifert et al. \cite{szeifert}, 
Wolf et al. \cite{wolf81}, \cite{wolf}, Walborn et al. \cite{walborn08}) depending on the phase of the LBV 
excursion, preventing  the determination of  specific line-ratio classification criteria. 

Furthermore, such Fe\,{\sc ii}-rich emission spectra are 
typically seen during and close to optical
{\em  maximum}, when LBVs are  thought to be at temperature {\em  minimum}, rather than in the hot phase supposed by Ma07 (cf. the above studies). 
Consequently, we 
prefer the historical term `Iron star' for such objects (cf. Walborn \& Fitzpatrick \cite{walborn00}) to simply reflect
the spectral appearance of the stars.

Similar issues are associated with the cool LBV classification. Specifically, the 
range of temperatures experienced by an LBV during an excursion appears to show an explicit luminosity dependance (e.g. Wolf \cite{wolf89}). For example
the high luminosity star R127 is observed to vary between  an Ofpe to a late B-/A-hypergiant classification (e.g. Walborn et al. \cite{walborn08})
whereas the low luminosity object HD~160529 only transits between B8 to A9 Ia$^+$ classifications (Stahl et al. \cite{stahl03}); 
thus HD~160529 would be classified by Ma07 as a cool LBV in both  hot {\em and} cool phases. 
Given that  the examples of this classification found within   M33  are  consistent with 
 B- and early A-hypergiants (e.g. Clark et al. \cite{clark10}, \cite{clark11b}, Walborn \& Fitzpatrick \cite{walborn90}),
 we  favour a simple morphological classification of blue hypergiant (BHG) for these stars.

Following this approach we suggest a  
classification of cool hypergiant for stars demonstrating the rich absorption+emission lines spectra associated  
 with  B324 and Var C in its cool phase (Sects. 3.5, 5.3.2 and  5.3.3). While a subset of such stars clearly may be LBVs undergoing excursions, some, such as B324 and the yellow hypergiants within Wd~1 (Clark et al. \cite{clark10}) show long term stability over decadal timescales, while others, such as $\rho$ Cas,
 instead transit to still cooler temperatures during outburst (Lobel et al. \cite{lobel03}).
While we recognise that this requires the re-assessment of the cool LBV candidates presented by Ma07,  as with the iron star and BHG   classifications, this 
terminology relies solely on spectral morphology and carries no explicit evolutionary  connotation, which may require additional multi-wavelength/epoch
 observations to determine.

\subsection{Iron stars}

We obtained spectra of four iron  stars (Table 1, Figs. 1 and A.1). Comparison  of these data to 
previous observations  revealed that none  of these stars unambiguously demonstrated behaviour
characteristic of the LBV phenomenon, although significant spectral variability was undoubtedly present in two of them. This is illustrated
by the spectra presented in Fig. 1, where for J013242.26+302114.1
 we see a clear change in the emission line ratio of the Balmer series in the $\sim$4~year period between observations,
 with a decrease (increase) in strength of H$\beta$ 
(H$\gamma$). Similar changes in the Balmer series over an identical time period  are also seen in J013324.62+302328.4, with a dramatic decrease 
in the strength  of H$\beta$ between the two observations.

A  number of spectra of J013406.63+304147.8 are present in the literature, dating from
from 1993-5 (Massey et al. \cite{massey06}), with subsequent spectra 
published by  Shemmer et al. (\cite{shemmer}), Sholukhova et al. (\cite{sholukhova}) and Fabrika et al. 
(\cite{fabrika}). While significant multiyear gaps exist between these observations, comparison of these to  
our new spectra  suggests that there has been a lack of conspicuous variability over this $\sim$17~yr period, with 
only minor  changes in the line profile of He\,{\sc i} 4471{\AA} and the anomalously strong Fe\,{\sc ii} 4556{\AA} 
line at a single epoch\footnote{Foreshadowing Sect. 5.2, 
 no characteristic LBV like excursions are apparent in the available photometry, although 
the  presence of a low level periodic photometric modulation (Shemmer et al. \cite{shemmer}) and 
spectroscopic radial velocity shifts lead Sholukhova et al. (\cite{sholukhova04}) to propose that J013406.63+304147.8 is a short 
period binary.}. 

Finally, despite a $>$15~yr baseline, we see no gross variation in spectral  morphology  between  the  $<$1995 spectrum of
Massey et al. (\cite{massey96}) and our  2003 Keck and 2010 WHT spectra of J013350.12+304126.6 beyond minor changes in emission line strength.

We also provide the first observations of three of these stars in the 6000-7000{\AA} window which encompasses H$\alpha$. In each case it is 
strongly in emission -  as expected given the behaviour of the higher Balmer transitions - with a single peaked non-P Cygni profile 
demonstrating strong emission wings. Nebular [N\,{\sc ii}] and [S\,{\sc ii}] is present in two of the four stars; given the 
comparative strength of the latter relative to the former 
we suspect this arises from a spatially coincident H\,{\sc ii} region rather than  CNO  processed
circumstellar ejecta. Remaining emission lines may predominantly be attributed to He\,{\sc i} and Fe\,{\sc ii}.

\subsection{B hypergiants}
We have made new spectroscopic observations  of five BHGs from Ma07 (Table 1, Figs. 1 and A.2). 
Historically, a classical temperature diagnostic for such stars is the He\,{\sc i} 4471{\AA}:Mg\,{\sc ii} 4481{\AA} line ratio 
(Walborn \& Fitzpatrick \cite{walborn90}), with the Mg\,{\sc ii} absorption strengthening relative to He\,{\sc i} for later spectral types.
Comparison of our spectra to those of Ma07 suggests that this ratio has varied in two  of the five stars examined;
J013416.44+303120.8 (illustrated in Fig. 1) and J013422.91+304411.0. Utilising this diagnostic, we would 
estimate {\em formal} spectral types of $\sim$B8 (2006) and $\sim$B2.5 (2010) for  J013416.44+303120.8 
(aided by comparison to examples presented in Monteverde et al. \cite{monteverde}) and $\sim$A0 (2006) and $\sim$B5-8 (2010) for 
J013422.91+304411.0, with the absence of He\,{\sc i} absorption in 
the spectrum presented by Ma07 suggesting a later spectral type in 2006 than 2010 (e.g. by comparison to IFM-B 1330; Monteverde et al. 
\cite{monteverde}).

However,  we caution against concluding that  a secular evolution in temperature - as might be expected for an LBV excursion - has occured
on this basis alone. Firstly, we note
that B-supergiants are prone to short period pulsational instabilities, as illustrated by examples found with Westerlund 1 (Clark et al. \cite{clark10}; see also Sect. 4) and we cannot rule such an explanation out on the basis of the current limited dataset. 
Moreover detailed quantitative non-LTE model atmosphere analysis of Galactic B-hypergiants by Clark et al. (\cite{clark11b}) found  that the behaviour of 
this line ratio may also depend on the location of the transition region between photosphere and wind; thus it does {\em not} serve as an unambiguous temperature diagnostic. Unfortunately, the low S/N ratio of the available spectra precludes us employing more robust diagnostics such as 
the Si\,{\sc iii}/Si\,{\sc ii} ratio.

Line profile variability (LPV) is also observed in the Balmer transitions of these stars. This is particularly notable in the H$\beta$ emission line of 
J013357.73+301714.2 (a weakening of the P Cygni profile from 2006), J013422.91+304411.0 (transition of P Cygni to pure emission
profile), J103424.78+303306.6 (weak P Cygni emission transitioning to absorption) and J013429.64+303732.1 (the evolution of a single
 peaked to a P Cygni profile superimposed on a broad base), while variability is also observed in the higher transitions of 
J013357.73+301714.2, J013416.44+303120.8 and J013429.64+303732.1. Comparable LPV has been observed in the
 (wind) emission  lines of Galactic B  super-/hyper-giants (e.g. Clark et al. \cite{clark10}, Morel et al. \cite{morel}, Kaufer 
et al. \cite{kaufer}) and LBVs  (Stahl et al. \cite{stahl03}). In the former it occurs over short ($\sim$days) timescales 
and has been associated with stochastic wind structure, while in the latter it is associated with the LBV cycle; the current lack of 
sampling precludes us  from unambiguously distinguishing between either possibility, although the lack of other associated spectral 
changes  argues against an LBV identification.

Unfortunately, the comparatively low S/N and resolution of the 
H$\alpha$ profiles presented in Ma07 precludes comparison to our data but we note that all five stars present an identical  profile 
consisting of  a narrow emission line superimposed on a broad base (Fig. A.2), 
similar to  those observed  in the mid- to late-B hypergiants in Westerlund 1 (Clark et al. \cite{clark10}).
Finally nebular [N\,{\sc ii}] and [S\,{\sc ii}] emission of ambiguous origin is present in four of the five stars observed although, as with the iron stars,
 we suspect this is not associated with circumstellar ejecta.

\subsection{P Cygni stars}

Ma07 identified four counterparts to the Galactic LBV P Cygni within M33, with spectra dominated by narrow emission profiles in the 
H\,{\sc i} and He\,{\sc i} lines; we have made observations of three of these (Figs. 1, 2 and A.3).   As with the BHGs, 
LPV is present and observed in the H$\beta$ line of J013341.28+302237.2 and the He\,{\sc i} lines of J013416.07+303642.1\footnote{Potentially 
significant nebular contamination of J013416.07+303642.1, revealed during attempts to model this star, prevents us from drawing conclusions as to variability of  the Balmer 
series.}. Discovery and subsequent follow up spectra of both stars extend the baseline of observations to $\geq 17$~yr for the former 
(Monteverde et al. \cite{monteverde}, Ma07) and $\geq 18$~yr for the latter (Spiller 
\cite{spiller},
Corral \cite{corral} and  Kehrig et al. \cite{kehrig}), but provide no 
evidence for (LBV driven) secular evolution during these intervals. For completeness we note that Urbaneja et al. (\cite{urbaneja05}) 
prefer a classification of B1 Ia$^+$ for J013341.28+302237.2 (=[HS80] 110A) which they find to be a highly luminous object with a 
high mass loss rate in comparison to Galactic examples (Clark et al. \cite{clark11b}); we return to this point in Sect. 5.3.3. 

In contrast, comparison of the available dataset  of J013339.52+304540.5 indicates dramatic long term variability (Fig. 2). Crowther et al. 
(\cite{b517}) first studied this star in detail, finding  that there was little evidence for photometric variability between the mid-1970s to 1990s and 
provided  spectroscopy from  1991, 1992 and 1995 which likewise indicated little or no evolution. Further low resolution spectra were subsequently 
obtained in 1992-4 (Corral et al. \cite{corral}), 1993 (Massey et al. \cite{massey96}) and 2006 (Ma07). Subject to the comparatively low 
resolution and S/N of the spectra, we find no evidence of variability between 1991-1992, but in 1993 the He\,{\sc i} 4471{\AA} line appeared to be  in 
absorption, before reverting to (P Cygni) emission by 1995. However, our new spectrum indicates a dramatic change in morphology in the intervening period,
 with the disappearance of He\,{\sc i}, He\,{\sc ii} and N\,{\sc ii} emission, a substantial weakening of the Balmer lines and the development of 
 weak photospheric features in e.g. Si\,{\sc iii}. The spectrum presented by  Ma07  reveals that the evolution had commenced prior to 2006, demonstrating 
an absence of emission from  high excitation species, albeit accompanied by significantly stronger emission in the Balmer lines than observed in either 
preceding or subsequent spectra.

Model atmosphere analysis of J013339.52+304540.5 by Crowther et al. (\cite{b517}) suggests 
$T_{\rm eff}\sim23$~kK, \.{M}${\sim}4.9\times10^{-5}M_{\odot}$~yr$^{-1}$ and $v_{\infty}\sim 275$~kms$^{-1}$ at this time (with the authors suggesting the 
 H/He$\sim$0.8 ratio was indicative of a post LBV phase). We may qualitatively attribute the change in spectral morphology to
 a decrease in both stellar temperature and wind density, the former indicated by the disappearance
 of high excitation features such as He\,{\sc ii} and the latter by the weakening of the wind emission lines.
 In principle one might hope to obtain an estimate of the temperature from the photospheric lines present. In comparison to J013341.28+302237.2 
($T_{\rm eff}{\sim}21$~kK), the lack of a strong C\,{\sc iii}+O\,{\sc ii} complex at $\sim$4650{\AA}, as well as the presence of  
Si\,{\sc iii} 4552, 68 and 75~{\AA}  absorption is suggestive of  a later classification  ($\sim$B2; $T_{\rm eff}{\sim}18$~kK), although we cannot exclude a 
somewhat earlier classification if the star is C-depleted (noting that subject to the caveats in Sect. 3.2, 
the absence of Mg\,{\sc ii} 4481{\AA} argues against a cooler temperature). Unfortunately, the 
current spectra are of insufficient S/N to be able to utilise the Si\,{\sc ii} and {\sc iv} transitions 
to obtain a more accurate temperature estimate; consequently  we are currently 
obtaining suitable spectra (and contemporaneous photometry) to permit such a detailed model-atmosphere analysis.

\subsection{Ofpe/WNL (=WN9-11h) stars}

 Ofpe/WNL (=WN9-11h) is a well defined stellar classification, although we note that at the low temperature extreme such stars share signficant morphological 
similarities with the P Cygni stars. This is well illustrated by a comparison of the spectra of J013416.07+303642.1
 (P Cygni type; Fig. A.3) 
and Romano's star (Ofpe/WNL type; Fig. A.4), both with one another and also with a (degraded) spectrum of P Cygni itself; we return to this topic in Sect. 5.3.1. 
Of the 12 such stars curently identified within M33 (Ma07, Neugent \& Massey \cite{neugent}) we have made observations of three; J013309.14+304954.5, J013327.26+303909.1 and 
J013509.73+304157.3 (henceforth Romano's star). 

Of these, minor changes in the strength of the wind emission lines appear present in J013327.26+303909 (not shown for brevity); such variability appears common 
amongst Galactic WRs  with a  number of mechanisms proposed to explain this (e.g. binarity, pulsations, wind structure). Unfortunately, 
the spectrum of J013309.14+303954.5 (=UIT045) is of rather poor S/N; despite 
this it would appear that the blend of N\,{\sc ii}+He\,{\sc i}+He\,{\sc ii}  emission that is prominent between 4600-4700~{\AA} in the discovery spectrum ($\leq$1995:
 Massey et al. \cite{massey96}) is now absent. Following the preceding dicussion,  the spectrum now bears close resemblance to the P Cygni-type stars. 
Nevertheless, further observations of both both stars will be required to first validate and subsequently constrain the nature of the (putative) variability.

In contrast, Romano's star is a well known  photometric ($\Delta V \sim16-18.6$)
and spectroscopic variable, demonstrating a $\sim$WN8h spectrum in visual minimum and as least as late as WN11h  in visual maximum
(Maryeva \& Abolmasov \cite{maryeva},  Polcaro et al \cite{polcaro}, Sholukhova et al. \cite{sholukhova})\footnote{Limited low resolution 
spectroscopy from 1992,
 coincident with the absolute visual maximum (Szeifert \cite{sromano}) is {\em suggestive} of a somewhat later B-supergiant 
classification. These appear
{\em not} to be dominated by the Fe\,{\sc ii} emission that characterises e.g. AG Car, HR Car or R127 during visual maximum
(Stahl et al. \cite{stahl01}, Szeifert et al. \cite{szeifert03} and Walborn et al. \cite{walborn08}), with the apparent lack of low excitation 
metallic photospheric lines suggesting an earlier classification than the F-supergiant spectrum demonstrated by e.g. M33 Var B (Ma07).}. 
Indeed following these recent studies it appears clear that it should be considered a {\em bona fide} LBV, a classification we adopt for the remainder of 
the paper.

Our spectrum implies a  $\sim$WN10h classification, compared to a $\sim$WN9h spectral type only 7~months earlier; 
consistent with the star simultaneously undergoing a secular photometric brightening. 
We  highlight the close similarity to P Cygni at this time; the sole observational differences  -  
the presence of weak He\,{\sc ii} 4686{\AA} and the lack of P Cygni profiles  in the Balmer series - consistent with
a somewhat higher temperature (Sect. 5.3.1).

\subsection{Bona fide LBVs}
 Finally, we turn to the three
 remaining stars - J013416.10-303344.9 and  M33 Var B and C -  which Ma07 denote
 as  LBVs (inevitably without spectroscopic classification criteria due to their variability;
Table 1 and  Fig. A.5). The 4000-5200{\AA} spectra of  J013416.10+303344.9 and M33 Var C obtained in 2010 and the 2003 spectrum  of M33 Var  B share very 
similar  
morphologies, being dominated by P Cygni Balmer and Fe\,{\sc ii} emission lines, which are strongest in the latter star. P Cygni He\,{\sc 
i} emission lines are present 
in M33 Var B and C, but are in absorption in J013416.10+303344.9.
The spectral morphology of  M33 Var B and C,  when observed in 2003 and 2010 (respectively), appear to resemble composites of the P Cygni and iron star
classifications (Fig. A.5), suggesting that the   physical properties of these stars
at the times of these observations were intermediate between these subtypes. Indeed, the spectrum of M33 Var B is strongly reminiscent of 
P Cygni (Fig. A.5)  - although the presence of Fe\,{\sc ii} emission  implies  a cooler, denser wind than that star.

In  contrast, the 2003 spectrum of M33 Var C is of a very different character, being dominated by a wealth of low excitation metallic absorption lines, 
 with only  H$\beta$ weakly in emission. Comparison to the 2010 observation and the highly luminous 
F0-5 Ia$^+$ hypergiant B324 (Fig. 2; Monteverde et al. \cite{monteverde}) indicates a significantly later spectral type at this epoch. While the H$\alpha$ line 
is saturated in the 2003 spectrum (as it is in the other 2 DEIMOS/Keck spectra), weak P Cygni emission is present in the 40 and 70 
multiplets of Fe\,{\sc ii}; again similar to B324 (Sect. 5.3.3).

Regarding variability, the {\em candidate} LBV J013416.10+303344.9 was observed by both Monteverde et al. (\cite{monteverde}) and 
Massey et al. (\cite{massey96}) between 1993-5; comparison of these data to our new spectrum show a broadly similar morphology, although the Balmer lines appear to have 
strengthened in this period. In contrast, the LBV classifications of M33 Var B and C are  long established (e.g. Szeifert et al. \cite{szeifert}, Viotti et al. 
\cite{viotti} and refs. therein). Ma07 present the 
most recent high resolution blue end spectra of M33 Var B that we are aware of (from 1993 and 2005); our spectrum differs notably from 
both of these, although in the absence of a contemporaneous lightcurve 
or better temporal spectroscopic  coverage we do not discuss this further. Finally, while clearly highly variable, 
 we defer analysis  of the spectral evolution of M33 Var C to Sect. 5.3.2, where it is discussed in the context of the long term photometric behaviour of the star.

\section{Photometry}

Photometric variability is a defining property of LBVs, and in addition to our new spectra, we may utilise existing photometric surveys (e.g. Hartman et
 al.\cite{hartman}, Shporer \& Mazeh \cite{shporer}, McQuinn et al. \cite{mcquinn}, Thompson et al. \cite{thompson})
of M33 to further constrain the nature of the candidate stars (Ma07, supplemented with objects listed in Table 2).

\subsection{Optical photomety}
Ma07 briefly reviewed the optical photometric data available for the candidate LBVs within M33, identifying 
variables both by comparison of historical to Local Group Galaxies Survey (LGGS; Massey et al. \cite{massey06}) observations and also via the 
multi-epoch photometric surveys of Hartman et al. (\cite{hartman}) and Shporer  \& Mazeh (\cite{shporer}). 
Expanding on this, we retrieved and  examined the individual lightcurves of the 15 stars flagged as variables by these  studies. Of these, only  one star
 demonstrated previously   unidentified, coherent long term variability - the iron star J013459.47+303701.0 (Fig. 3). While this is potentially consistent 
with an LBV identification we note that the star became bluer as it brightened, whereas the opposite behaviour might be expected, with LBVs typically 
cooling as they brighten.

Of the known LBVs, the 2000-2003 lightcurve of M33 Var C (Shporer \& Mazeh \cite{shporer}) is of particular interest (Fig. 4), since it indicates 
a $\sim$0.8~mag. brightening between  2001-2 July, followed by an apparent plateau with $ V \sim 15.4$~mag (comparable to the maximum light of the two previous maxima; Humphreys et al. 
\cite{humphreys88}).
We return to discuss the combined photometric  and spectroscopic dataset for this star in Sect. 5.3.2.

After excluding  known (photometrically variable) LBVs such as M33 Var C from the sample,  apparently aperiodic photometric variability of $\leq$0.4mag. amplitude was  associated with the remaining objects (e.g. J013426.11+303424.7; Fig. 2)\footnote{The iron stars
J013235.25+303017.6, J013242.26+302114.1, J013300.02+303332.4, J013406.63+304147.8 and J013426.11+303424.7, the Ofpe/WN9 stars  
J013309.14+304954.4, J013327.26+303909.1 
and  J013353.60+303851.6, the B hypergiants J013355.96+304530.6 and J013416.44+303120.8 and the P Cygni candidate J013416.07+303642.1.}. Likely driven 
by pulsations, 
quasi-periodic photometric variability 
over timescales of tens to hundreds of days  appears ubiquitous amongst supergiants of spectral type B and latter  - the so called $\alpha$ Cygni variables (van 
Leeuwen et al.
 \cite{alpha}, Clark et al. \cite{clark10})   - as well as LBVs (Lamers et al. \cite{lamerslbv}) and it is tempting to attribute the variability to this cause (with a 
combination of changes in the pulsational period and poor temporal sampling accounting for the lack of identifiable periodicity). Finally, we note that
 despite rather large amplitude ($\Delta V \sim0.8$mag.) historical photometric and spectroscopic variability (Ma07; Fig 1), 
the iron star J013242.26+302114.1 was not flagged as variable by either survey, nor were a number of the spectroscopic variables identified in this 
work\footnote{J013324.62+302328.4, J013341.28+302237.2, J013422.91+304411.0, J013357.73+301714.2, J0134234.78+303306.6 and J013429.64+303732.1},
including the strong LBV candidate J013339.52+304540.5 (Sect 3.3 and Fig. 2; 
although pre-empting the following section, the BHG J013429.64+303732.1 is a mid-IR variable). 

We may also utilise the LGGS photometry to construct colour manitude plots for the candidates of Ma07, which we show in Fig. 5. We can clearly identify the 
 blue plume of highly luminous blue supergiants that Massey et al (\cite{massey06}) describe. However, plotting both $(B-V)$ and $(V-I)$ colour indices demonstrates that a number of 
individual objects such J013340.60+304137.1 show rather discrepant colours, unfortunately making dereddening individual stars problematic.

Nevertheless, of the objects plotted, the brightest, J013350.92+303936.9 (=UIT 218, which Ma07 label as a hot LBV candidate) is in fact the spatially extended nucleus of M33, which
Massey et al. (\cite{massey96}) show has an F supergiant like spectrum (and hence we do not consider it further). The next brightest, J013355.96+304530.6 
(=B324) appears to be both intrinsically bright and possesses the latest spectral type  of any object considered here - we return to this 
in Sects. 4.2 and  5.3.3. 

Of the remaining objects from  Ma07, the Ofpe/WNL stars 
appear to be the  faintest subtype, with $V >17.5$, although there is some overlap  with both the  P Cygni and 
B hypergiants. With $V <18$ and  $<17.5$ respectively, these subtypes appear systematically  brighter (as might 
be expected from  the larger visual bolometric corrections anticipated for  the Ofpe/WNLs).
Given their intrinsic variability, {\em bona fide}  LBVs span a wide range of magnitudes, 
with  Romano's star 
having been  found to be at  least 0.6~mag fainter  in the V band than shown here  (e.g. Sholukhova et al. \cite{sholukhova}). 
Finally, and somewhat unexpectedly, the iron stars are found
to span the largest magnitude range  ($\Delta$V $\sim$3.5~mag); we return to the nature of these objects in Sect. 5.2.

\begin{figure}
\includegraphics[width=5cm,height=9cm,angle=270]{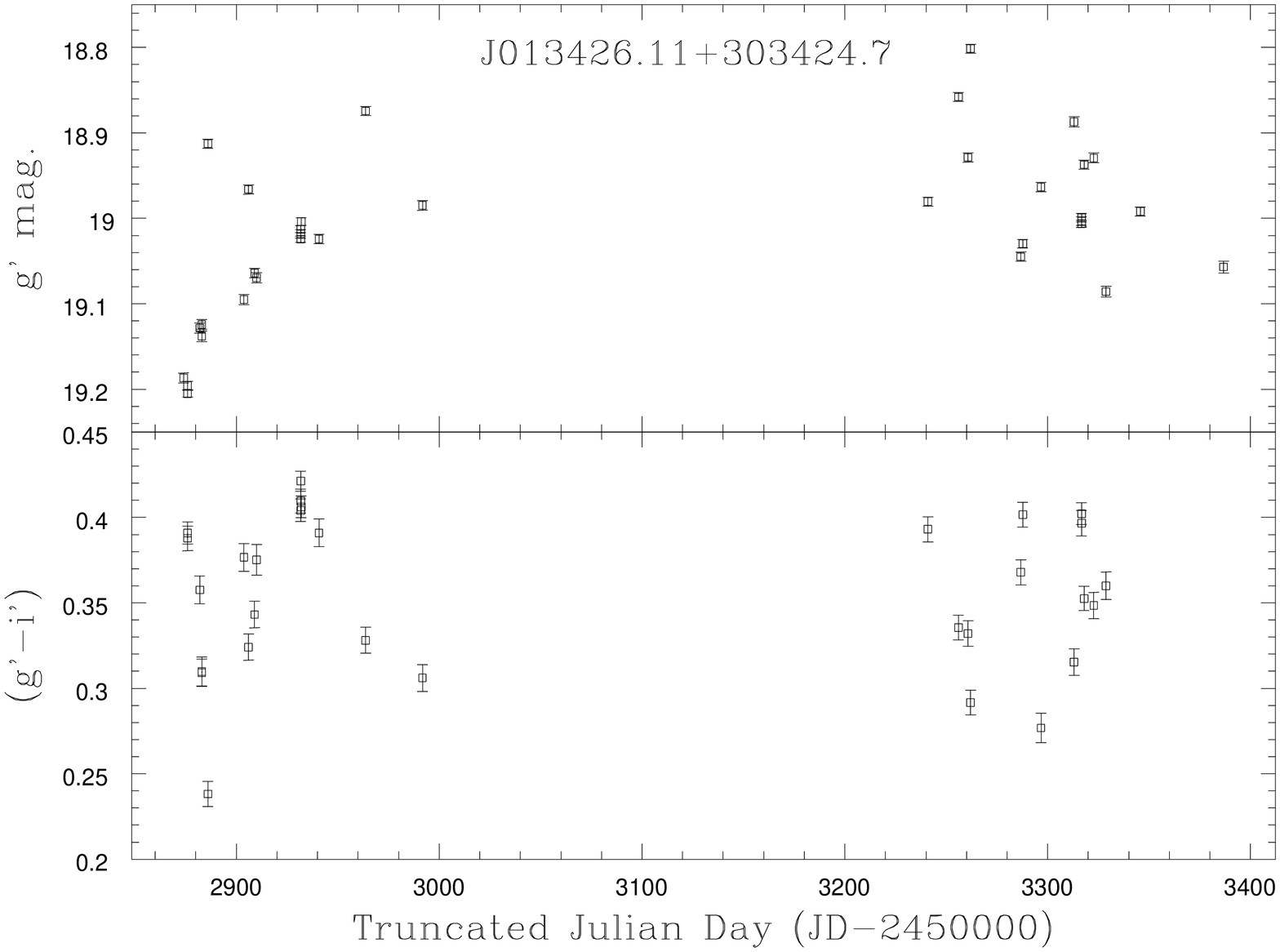}
\includegraphics[width=5cm,height=9cm,angle=270]{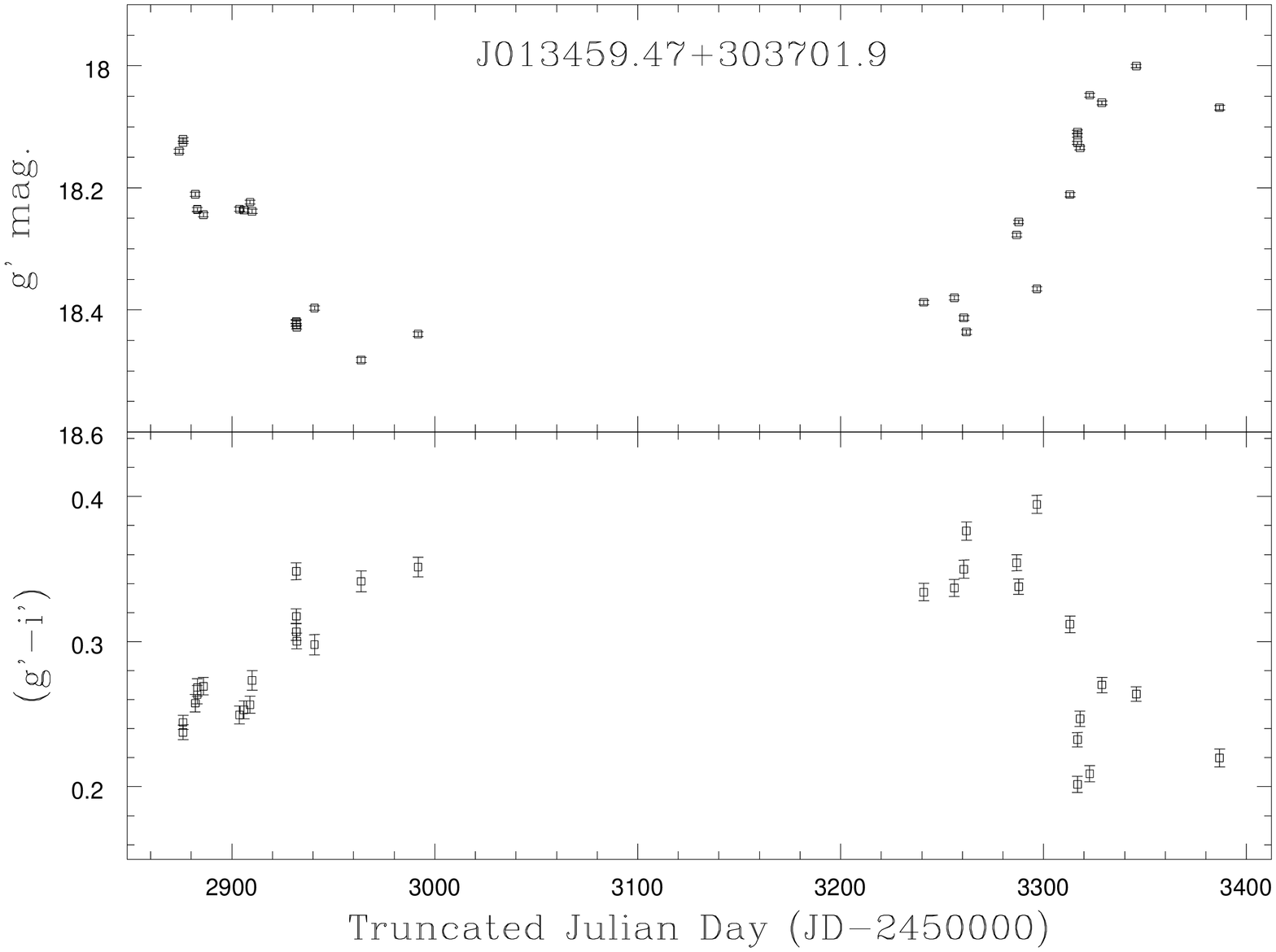}
\caption{Sloan g' and (g'-i') colour index plotted against time for the
candidate LBVs  J013426.11+303424.7 and J013459.47+303701.0. Data from Hartman
et al. (\cite{hartman}).}
\end{figure}

\begin{figure}
\includegraphics[width=5cm,height=9cm,angle=270]{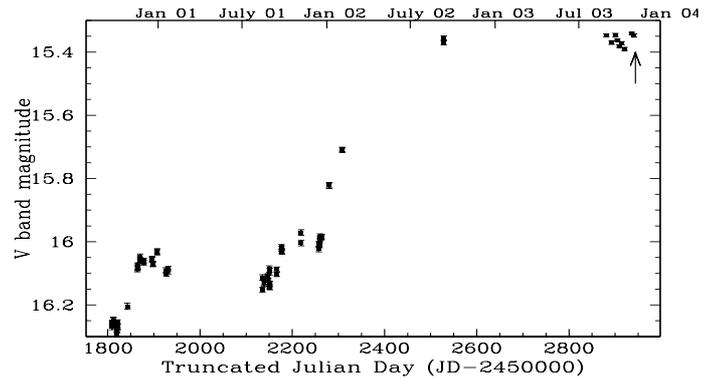}
\caption{$V$ band photometry of M33 Var C from Shporer \& Mazeh (\cite{shporer}),
with the timing of the spectroscopic observation discussed in Sect. 3.5 
indicated.}
\end{figure}

\begin{figure*}
\includegraphics[width=8cm,angle=0]{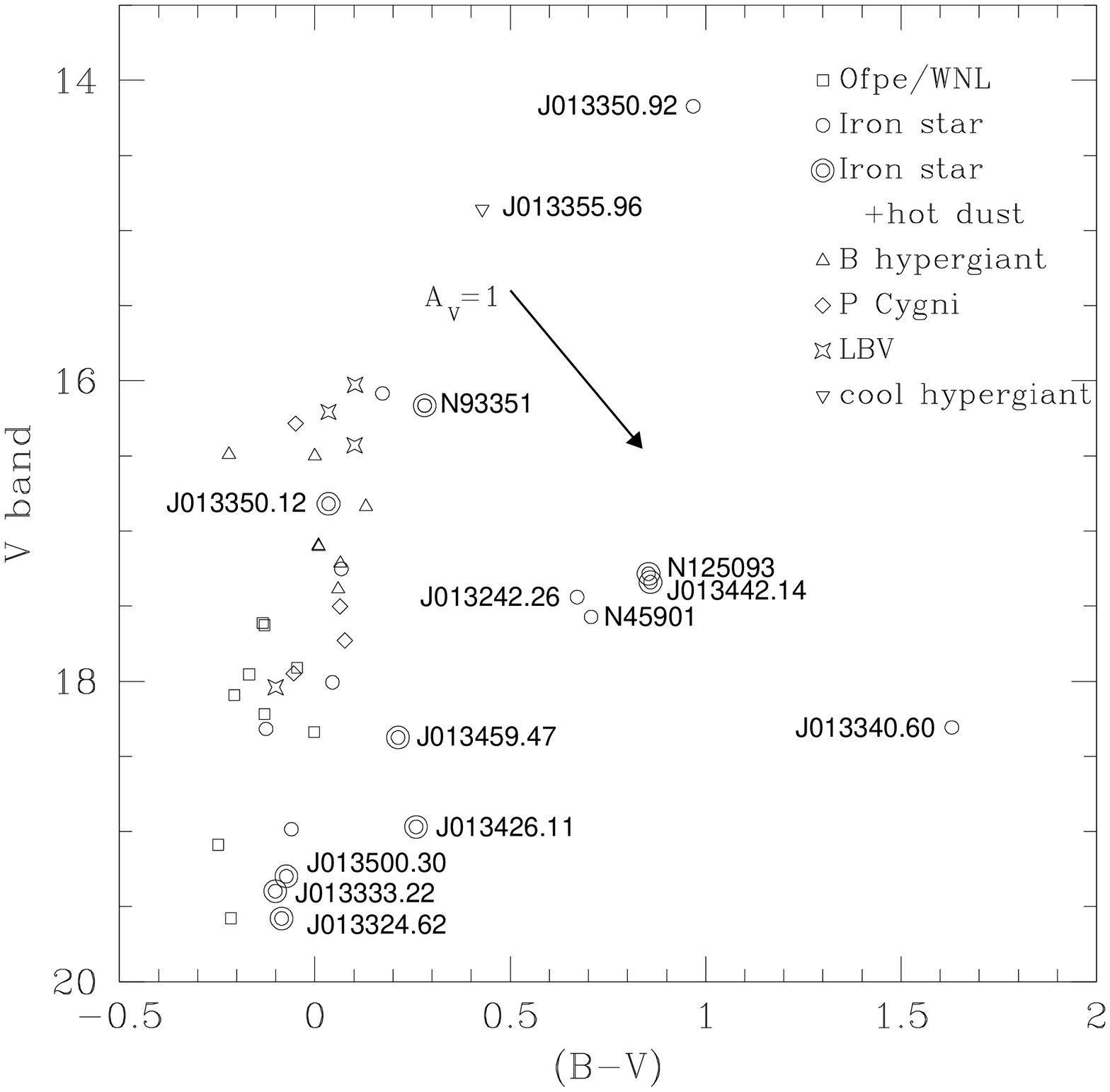}
\includegraphics[width=8cm,angle=0]{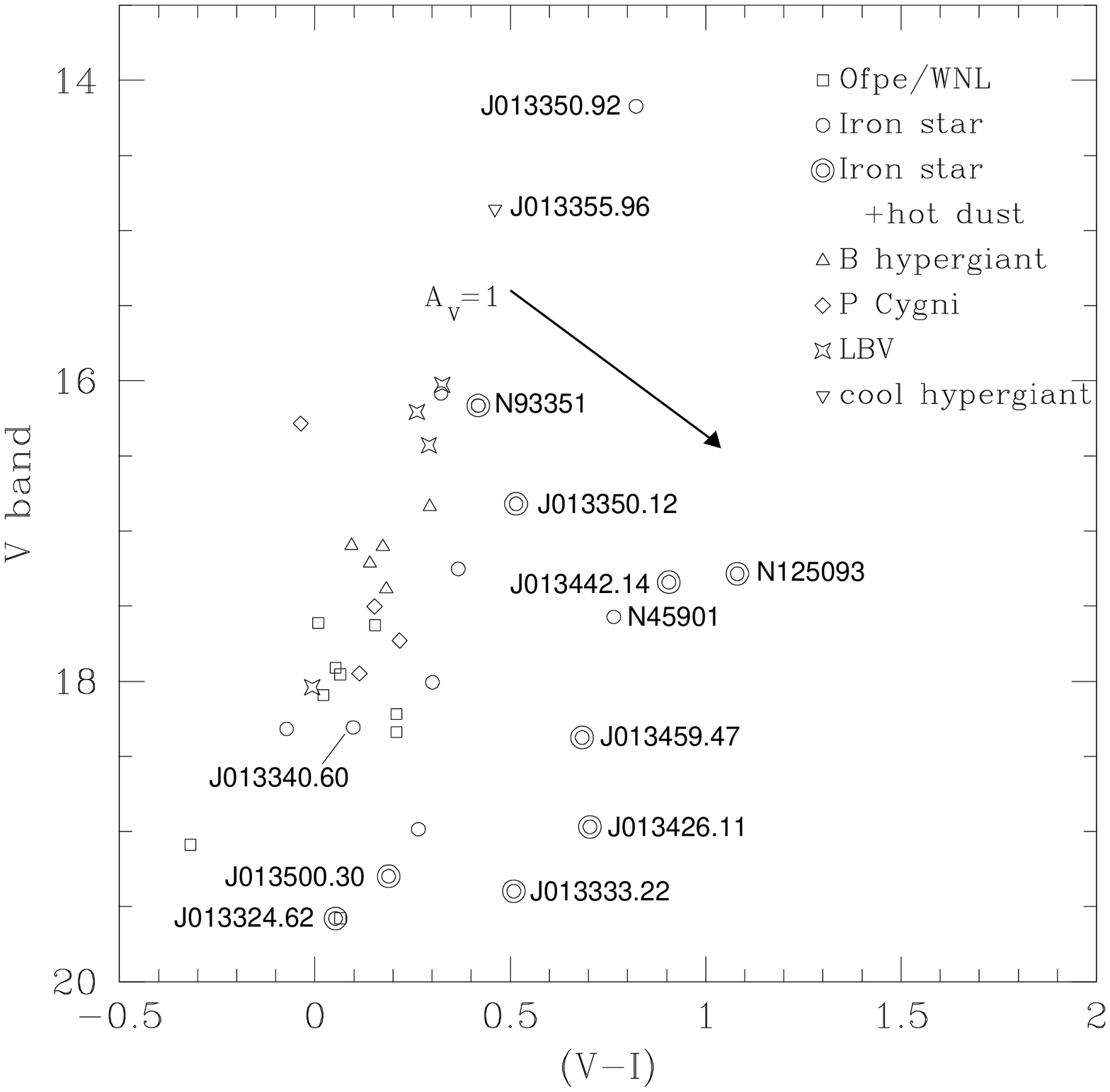}
\caption{Colour magnitude plots for (candidate) LBVs within M33, following 
the ammended classification scheme of Ma07. Stars associated with hot dust are 
identified in Sect. 4.2. Note that Massey et al. (\cite{massey96}) associate the brightest object,  J013350.92+303936.9 (=UIT218), 
with the galactic nucleus - an apparently composite source with  an F Ia spectral type rather the hot LBV 
classification given in Ma07 - while J013242.26-301214.1 lacks an $(R-I)$ indice.}
\end{figure*}

\subsection{Mid-IR photometry}

\begin{figure*}
\resizebox{\hsize}{!}{\includegraphics[width=10cm,angle=0]{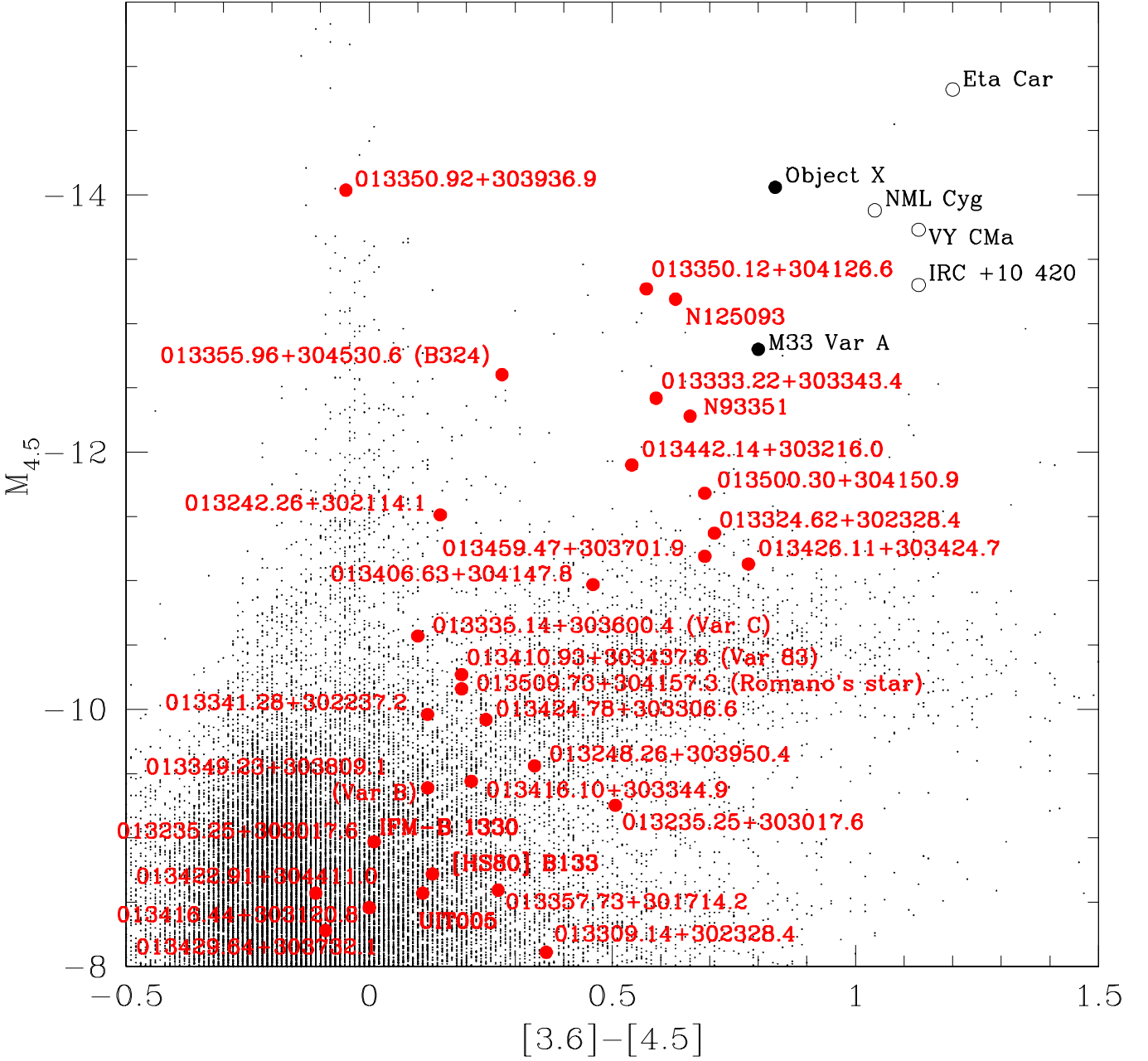}}
\caption{Mid-IR colour magnitude plot of stellar point sources within M33. Related 
dusty, high  mass loss rate massive evolved objects within M33 and the Galaxy 
 also plotted for comparison (filled and empty black circles respectively). See Sect. 4.2 for origin of individual data points.
Note that J013350.92=303936.9 (=UIT218) is associated with the galactic nucleus and is likely a composite source.}
\end{figure*}

\begin{figure*}
\resizebox{\hsize}{!}{\includegraphics[width=10cm,angle=0]{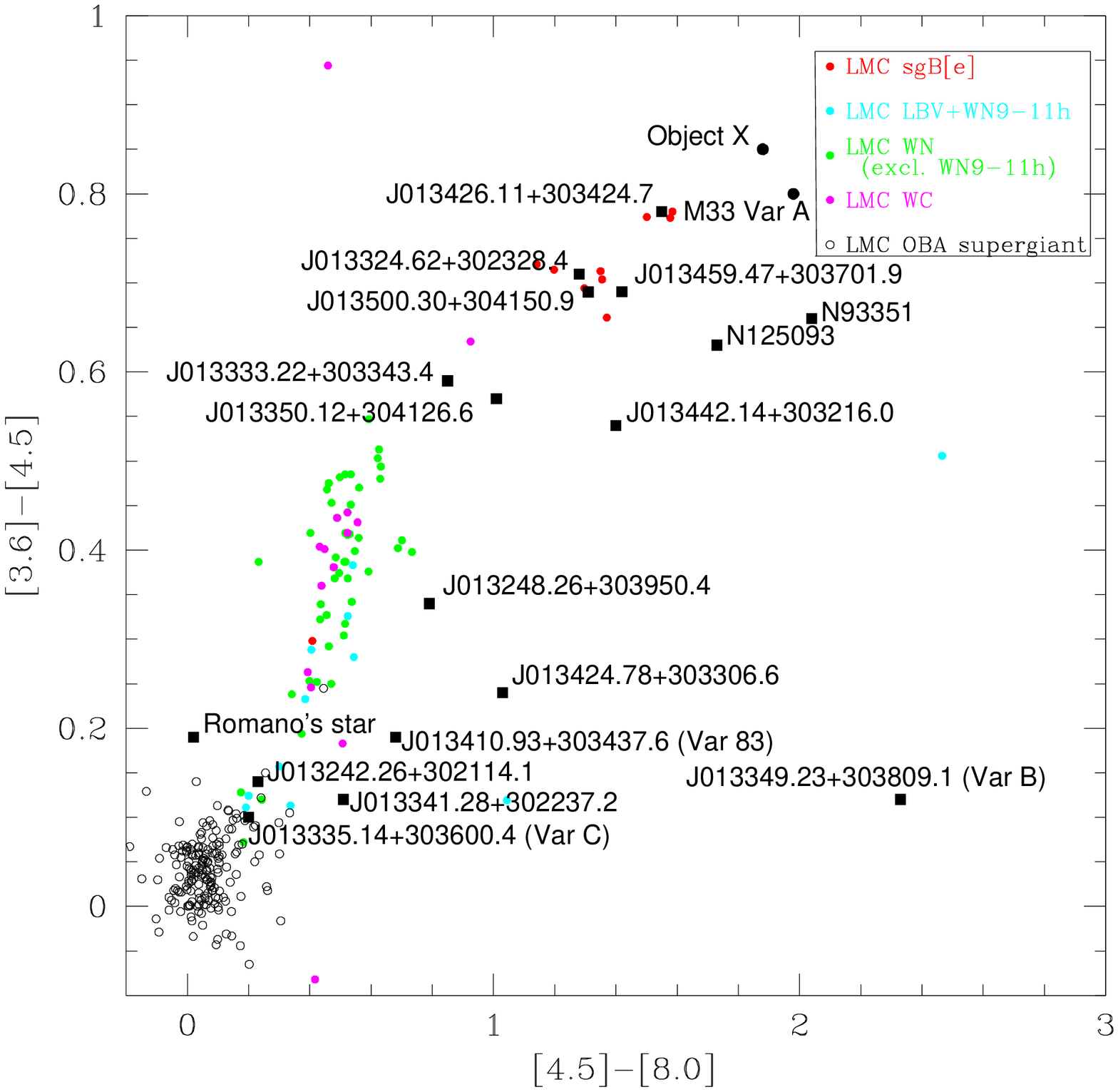}}
\caption{Mid-IR colour colour plot of candidate LBVs and related stars within M33, with  the massive stellar
 population of the Large Magellanic Cloud shown for comparison (data from Bonanos et al. \cite{bonanos}).
The outlying LMC WC star with [3.6]-[4.5]$\sim$0.94 is the heavily reddened triple system BAT99-38, while the 
LMC LBV with [4.5]-[8.0]$\sim$2.5 is R71.}
\end{figure*}

M33 has  been the subject of extensive mid-IR studies, with six epochs of {\em Spitzer}/IRAC  (3.6-8.0$\mu$m) observations between
2004 January and 2006 February  (unfortunately not contemporaneous with  the LGGS photometry or the  spectroscopy of Ma07 and this work). Point source catalogues 
derived from these data have been compiled by McQuinn et al. (\cite{mcquinn}) and  Thompson et al. (\cite{thompson}) 
 enabling identification with the source list of Ma07; a task undertaken by the latter authors. To this list we were able to add the  
additional candidates listed 
in  Table 2 and cross correlated the expanded census against both catalogues (as well as adopting fluxes for four sources kindly 
provided by  Valeev 2011, priv. comm.).

We found a total of 29 sources had a detection at one or more wavebands. Of these 18 were from the catalogue of Thompson et al. 
(\cite{thompson}; including three BHGs previously unconsidered by these authors), four supplied by 
Valeev et al. (\cite{valeev09}) and seven new counterparts  unique to McQuinn et al. (\cite{mcquinn}). 
Of the 16 stars from the catalogue of Ma07 identified as having mid-IR counterparts by Thompson et al. (\cite{thompson}), 13 
were found in the  catalogue of McQuinn et al. (\cite{mcquinn}).
 These showed systematic offsets of m$_{3.5}\sim$0.15~mag and  m$_{4.5}\sim$0.2~mag  in the sense that the magnitudes in 
Thompson et al. 
(\cite{thompson}) were brighter; a corresponding   correction has {\em not} been applied to the values presented in Table 3 but {\em has} been applied
 in Figs. 6 and 7 to enable direct comparison. No obvious offset was found with respect to 8$\mu$m fluxes, although we caution  that individual  sources in all 
3 
bands show significantly greater offsets than described here\footnote{J013333.22+303343.4 at 8$\mu$m, Var B at 
4.5$\mu$m, J013350.92+303936.9 at 3.6$\mu$m and J013416.10+303344.9 at 4.5$\mu$m}; in such cases we adopt the 
relevant fluxes from Thompson et al. (\cite{thompson}).

Broken down by nature of source\footnote{Excluding the nuclear cluster  source J013350.12+304126.6, reclassifying Romano's star as a 
{\em bona fide} LBV and  adopting a classification of iron star for N45901, N93351 and N125093 (Table 2).}    14 of 18 iron stars, 9 of the  13 BHGs, one each of the 11 Ofpe/WNL and four P Cygni candidates, all four LBVs and the sole cool hypergiant, B324, are detected in one or more wavebands. Fluxes for those sources not listed by Thomspon et al. (\cite{thompson}; their Table 3)
 are given in Table 3 of this paper. Only three stars are found to be variable;  M33 Var C, J013429.64+303732.1 (Thompson et al. \cite{thompson}) and Romano's star.
The lightcurve of M33 Var C (Fig. A.3 of Thompson et al. \cite{thompson}) is particularly interesting, since fortuitously the Spitzer observations commenced at the 
point  when the  optical observations of Shporer \& Mazeh (\cite{shporer}; Fig. 4) ended and indicate a $\sim$0.8mag fade over a period of 
$\sim$400~days commencing early in 2005; we return to this in Sect. 5.3.2.

A number of  different physical processes might be expected to contribute to the $\sim$3-8$\mu$m fluxes of the sources, including  photospheric emission (potentially 
 from a putative cool companion), free-free continuum emission from a dense wind and  thermal emission from a dusty circumstellar environment (either resulting 
from dust condensation in a wind shock in a colliding wind binary or in red supergiant/LBV ejecta). Given this, the current uncertainty in the parameters of the underlying stars
(and hence the stellar contribution to the composite spectral energy distribution) and the potential difficulty   in accounting for the contribution of 
 line and molecular (e.g. PAH) emission in the continuum bands, quantitative analysis of individual sources appears
premature. Nevertheless, we may attempt to determine the likely nature of the mid-IR emission from different objects in a statistical sense utilising various 
different mid-IR colour/magnitude and colour/colour diagnostic plots that  have been developed for such evolved massive stars (e.g. Hadfield et al. \cite{hadfield}, Bonanos et al. 
\cite{bonanos}).

Following Thompson et al.  we initially utilise the [3.6]$-$[4.5] vrs. M$_{4.5}$ colour/absolute magnitude plot 
(Fig. 6 -  employing a distance modulus, $\mu$=24.92; Bonanos et al. \cite{bonano06}).
The stellar  population of  M33 is predominantly located  in the lower left of the plot, 
with a plume extending to larger M$_{4.5}$, that  likely represents a combination of foreground 
and intrinsically luminous (composite) sources. A population of  dusty, (extreme) AGB stars appear to dominate the region 
of the diagram delineated by M$_{4.5}>-11$ and approximately [3.6]-[4.5]$>$0.5 (Thompson et al. \cite{thompson}).
Finally, a population of  massive,  evolved stars are located in the  upper right corner. Following these authors and 
Khan et al. (\cite{khan}), we plotted selected Galactic objects here for comparison, noting that the
[3.6]-[4.5] colour indices of these stars suggest the presence of hot dust 
(with a simple black body temperature, $T_{\rm BB}\sim 350-1000$~K). 

We may supplement this with the [3.6]$-$[4.5] vrs. [4.5]$-$[8.0] colour colour plot of Bonanos et al. (\cite{bonanos}; Fig. 7). 
The nature of the emission from sources in this diagram is best demonstrated by plotting the massive stellar population of the Large 
Magellanic Cloud (excluding the cool supergiants; Bonanos et al.). The OBA super-/hypergiant and WRs form 
a continuous  sequence of increasing mid-IR colour excess, with  the latter redder than the former. This may simply 
be  understood as the increasing contribution from free-free emission, which on average is more 
pronounced for the WRs due to their denser winds. LBVs and Ofpe/WNL stars are also found along this sequence, with an 
additional contribution from circumstellar dust explaining the outliers. Finally, stars exhibiting the B[e] phenomenon form a 
distinct group, with their very red colours the result of the hot dust that characterises such objects.

In the colour magnitude plot (Fig. 6), the majority (18) of the stars with mid-IR detections appear to be broadly 
 co-located with the underlying stellar population of M33 (approximately bounded by M$_{4.5}>-11$ and  [3.6]$-$[4.5]$<$0.5); these consist 
of representatives of 
each of the ammended morphological classifications of Ma07 (bar the single cool hypergiant, B324). 
In the absence of quantitative  modeling, the nature of the IR emission of these stars is uncertain, 
although we suspect it is due to a combination of wind (free-free) and photospheric emission. Indeed
the three mid-IR variables - M33 Var C,  Romano's star and  the BHG J013429.64+303216.0 - are located in 
this region and none show correlated mid-IR colour/magnitude changes (e.g. in the sense that they become redder as they 
brighten),  as might be expected if the emission originated in a variable dusty circumstellar environment.

Unfortunately, only seven of these stars have 8$\mu$m detections (Fig. 7). Of these,  Romano's star and M33 Var C 
are co-located along the OBA supergiant/WR  sequence; suggestive of wind+photospheric emission (although Valeev et al. 
\cite{valeev09} attributes the IR emission to hot dust, highlighting the difficulty in interpretation). The remaining  objects\footnote{
The LBVs M33 Var 83 and Var B, the iron star  J013248.26+303950.4, the BHG J013424.78+303306.6 and the P Cygni star
J013341.28+302237.2} 
are all displaced from this locus by virtue of their comparatively strong 8$\mu$m emission, implying the presence of warm    
circumstellar dust. Note however, that the absence  of a pronounced 8$\mu$m flux does not mean that the remaining stars 
lack dusty ejecta, since emission from LBV nebulae {\em typically} peaks at  significantly longer wavelengths 
(e.g. Egan et al. \cite{egan}, Clark et al. \cite{clark03}, Voors et al. \cite{voors}).
  
A further 11 stars are found outside this region of the colour magnitude plot. Of these, 
 the mid-IR colours of the iron star J013242.26+302114.1 are suggestive of  wind+photospheric emission 
(Fig. 7) and while no 8$\mu$m detection is available for J013355.96+304530.6 (=B324), we 
suspect a similar origin for its $3.6 - 4.5\mu$m flux, given both its intrinsic extreme luminosity and strong  wind emission 
lines (Sect. 5.3.3).

This then leaves nine mid-IR luminous stars (M$_{4.5}< -11$),  {\em all} of which are iron stars\footnote{Note  
that identification of the most luminous of these sources as M33 Var C in Thompson  
et al. (\cite{thompson}) is in error.}. These are found in a sequence of increasing
M$_{4.5}$ and [3.6]$-$[4.5] colour (Fig. 6) that terminates amongst highly luminous evolved stars with rich dusty circumstellar 
environments such as the Galactic RSGs  NML Cyg (Schuster et al. \cite{schuster}) 
and  VY CMa (Smith et al. \cite{smith01}), the  LBV $\eta$ Car (Smith et al. \cite{smith02}), the YHG IRC +10 420 (Bl{\"o}cker et al. \cite{blocker}) and, within 
M33, the cool transient M33 Var A (Humphreys et al. \cite{humphreys06}) and Object X (Khan et al.\cite{khan}). Of 
these, four  - 
 J013324.62+302328.4, J013426.11+303424.7,  J013459.47+303701.9 and J013500.30+304150.9 - are co-located with the sgB[e] stars (Fig. 7),
 while the colours of   a further two  - J013333.22+303343.4 and J013350.12+304126.6 - are also suggestive of (a smaller quantity of) hot dust. 
The final three stars - N93351, N125093 and J013442.14+303216.0 - also appear associated with hot dust, although their larger 8$\mu$m fluxes 
imply greater quantities of warm dust is also  present with respect to the former objects. Unfortunately, due to the lack of longer wavelength data 
 we refrain from  attempting to determine temperature(s) for the dust component(s)\footnote{Indeed, even with longer wavelength data, the  study of Object X by 
Khan et al. (\cite{khan}) illustrates the difficulty in quantitative analysis, presenting two acceptable fits to the observed spectral energy distribution
arising from the adoption of very different  temperatures and chemical  properties  of the circumstellar  dust.}, although  Bonanos
 et al. (\cite{bonanos}) report dust temperatures of $\sim$600K for those LMC sgB[e] stars for which 24$\mu$m fluxes are 
available.

These stars are found to span an unexpectedly wide range of $\sim$3 magnitudes  
in the $V$ band (Fig. 5), including objects that appear to be either  intrinsically faint (e.g. J013324.62+302328.4) 
and/or subject to significant reddening (e.g. J013459.47+303701.9). 
Similar behaviour is also found at mid-IR wavelengths, with the nine stars spanning $\geq$2.3 mags in the 4.5$\mu$m bandpass. The two brightest objects - 
J013350.12+304126.6 and N125093 - are comparable to the cool transient M33 Var A and the Galactic YHG IRC +10 420 (Fig. 6). However their mid-IR colours
suggest differences in dust temperature in comparison to these objects, particularly for J013350.12+304126.6 which, by virtue of its 8$\mu$m flux, 
 appears to lack a warm dust component.  

Turning to the fainter extreme and we find that the four faintest mid-IR sources from this subset are those that have mid-IR
 colours consistent with sgB[e] stars. With $V>18$~mag. these are also amongst the faintest stars at optical wavelengths (Fig. 5). 
Bonanos et al. (\cite{bonanos}) report that the LMC sgB[e] stars experience greater  reddening  than 
 normal OB stars; given the reddened optical colours of J013426.11+303424.7 and J013459.47+303701.9 we suspect that this may also be the case for these stars.
An (extreme) example of this behaviour would then be the evolved massive star 
`Object X',  where extensive  recent mass loss has rendered it the brightest stellar object within M33 in the  mid-IR,
 but undetectable at wavelengths  shortwards of the V band (Khan et al. \cite{khan}). 
Applying magnitude (M$_{4.5}<-11.5$) and colour (0.5$<$[3.6]$-$[4.5]$<$1.5) cuts designed to identify massive 
dusty objects, Thompson et al. (\cite{thompson})  identified a further 16 optically faint and 9 optically undetectable
 sources within M33, and it is tempting to attribute our subset of optically faint iron stars  associated with hot dust
  to a low extinction tail of this population; we return to the nature of these objects in Sect. 5.2.

\begin{table*}
\begin{center}
\caption[]{Mid-IR fluxes for new candidate LBVs and BHGs}
\begin{tabular}{lccccc}
\hline
\hline
Name & Class. & [3.6] & [4.5] & [5.8] & [8.0]  \\
\hline
UIT005  & BHG & 16.46 & 16.35 & -& - \\
 {[}HS80] B133 & BHG & 16.33 & 16.20 & -& - \\
IFM-B 1330 & BHG & 15.96 & 15.95  & -& -\\
\hline
M33 Var B & LBV & 15.65 &15.53& 14.82 &  13.20\\
N93351  & iron star &13.30 & 12.64 &11.67  &  10.60 \\
N125093 & iron star & 12.36 & 11.73 & 11.05 & 10.00 \\
Romano's star & LBV &14.95 & 14.76 & 14.72 & 14.74\\
\hline
013235.25+303017.6 & iron star & 16.32 & 15.87 & - & - \\
013242.26+302114.1 & iron star &13.70 & 13.61 & - & 13.18 \\
013309.14+302328.4 & Ofpe/WN9 &17.32 & 17.01 & - & - \\
013350.92+303936.9 & iron star  &10.99 & 11.08 & - & 11.40 \\
013355.96+304530.6 (=B324) & YHG? & 12.74 & 12.52 & - & - \\
013357.73+301714.2 & BHG &16.74 & 16.53 & - & - \\
013416.44+303120.8 & BHG & 16.61 & 16.66 & - & - \\  
\hline
\end{tabular}
\end{center}
{{\em Top panel:} fluxes  from
the catalogue of Thompson et al. (\cite{thompson}), {\em middle panel:} fluxes  from Valeev et al. (\cite{valeev09}, 
priv. comm. 2011) and {\em bottom panel:} fluxes from the McQuinn et al. (\cite{mcquinn}).
Note that the values from McQuinn et al. (\cite{mcquinn}) have {\em not} been subject to correction due to a 
systematic offset with those of Thompson et al. (\cite{thompson}), while the former do not provide [5.8] magnitudes. Classifications given follow the ammended  morphological
scheme of Ma07, noting  that Massey et al. (\cite{massey96}) associate  013350.92+303936.9 (=UIT218) with the galactic nucleus; hence it  is likely that it is a composite source.}\\
\end{table*}

\section{Discussion}

A central goal of this work is the determination of  the physical and evolutionary nature of the stellar sample.  It is easy to place both the 
Ofpe/WNL and P Cygni  stars at the high temperature extreme of an LBV excursion, with the former appearing to 
be both hotter and to have larger wind terminal velocities (for a comparable mass loss rate, leading to a lower wind density) than the latter 
(e.g. Clark et al. \cite{liege} and in prep.). However, how may  the remaining stars be accommodated?

\subsection{The BHGs}

  Clark et al. (\cite{clark11b}) studied the Galactic population of BHGs and found that they appeared to result from two distinct channels, with high 
luminosity examples such as $\zeta^1$ Sco  found in a pre-LBV phase and lower luminosity objects such as HD~160529 being post-RSG objects potentially 
encountering the LBV phenomenon as they evolve to higher temperatures.  Given the concerns  regarding both dereddening individual stars and  the use of the He\,{\sc i} 
4471/Mg\,{\sc ii} 4481 {\AA} ratio as a temperature diagnostic (Sect. 3.2) we refrain from calculating  bolometric luminosities for individual BHGs discussed
in this study.  Nevertheless assuming negligible reddening, a bolometric correction appropriate for an A0 spectral classification (BC$_{\rm A0}$=-0.15mag and for 
reference BC$_{\rm B2}$=-1.59mag; Clark et al. \cite{clark05b})   and a lower limit of V$\sim$17.5 for the range of optical magnitudes spanned by 
these stars, results in  a {\em minimum} bolometric luminosity of log(L/L$_{\odot}$)$\sim$5.0; consistent with both low and high luminosity/mass channels.
Thus their individual placement in an evolutionary scheme will have to await systematic model atmosphere analysis, although of the three BHGs subject to such a study 
 (Urbaneja et al. \cite{urbaneja05}, \cite{urbaneja11}) [HS80] 110-A is found to be unexpectedly luminous, (Sect. 5.3.3), while the  
stellar+wind properties of  both UIT 005 and OB10-10 are similar to those of pre-LBV  Galactic BHGs (Clark et al. \cite{clark11b}).

\subsection{The iron stars}

\subsubsection{LBV candidates}

While resulting in a heterogenous group of stars (Sect. 3), the observational criteria employed by Ma07 to define the class of iron stars (Sect. 3.1) are clearly 
successful in identifying LBV candidates. The candidatures of N45901, N93351 and N125093 are discussed by Valeev et al. (\cite{valeev09}, \cite{valeev10}), and
are supported by pronounced photometric variability (N93351 and potentially the remaining stars), mid-IR excesses (N93351 \& N125903; Sect. 4.2) and the presence
of [Ca\,{\sc ii}] ${\lambda}{\lambda}$ 7291, 7323 emission (N93351 and N125093), which appears associated with  variable/eruptive stars such as M33 Var A and
 IRC +10 420.  The impressive similarity of the spectrum of J013350.12+304126.6 to that of the LBV R127,  together with its 
 pronounced mid-IR excess, flags it as a strong candidate even in the absence of current variability, while the mid-IR properties of J013442.14+303216.0 also
points to the presence of a cool dusty circumstellar environment. J013459.47+303701.9 also appears dusty and 
demonstrates coherent photometric variability (Fig. 3); although it is optically fainter than the preceding examples  this may in part be due to 
greater (intrinsic) reddening. Finally  Ma07 report striking spectroscopic variability in J013332.64+304127.2 
(from a WNL to iron star morphology), although unlike the preceding examples this star does not appear to support an IR excess.

Therefore, under the assumption that a subset of both the iron stars and BHGs are indeed {\em bona fide} LBVs,  
consideration of confirmed high and low luminosity LBVs (e.g. AG Car, HR Car, R71, R127 and HD~160529) allows us to  infer  a broad
physical progression from high to low temperatures in terms of the ammended classification  scheme of Ma07: \newline

Ofpe/WNL $\rightarrow$ P Cygni $\rightarrow$ iron star/BHG $\rightarrow$ cool hypergiant \newline

 This sequence is subject to three  caveats; that (i) progress between subtypes is continuous (cf. the apparent hybrid spectra of M33 Var B; Sect. 3.5), 
(ii) the precise appearance of the spectra are also  subject to the properties of the stellar wind, which are also found to vary through an LBV excursion 
(e.g. Groh et al.\cite{groh}) and (iii) depending on stellar luminosity, the full range of subtypes may not be experienced,
 with lower luminosity examples not reaching the hottest Ofpe/WNL and P Cygni phases (cf. Wolf \cite{wolf89}).

\subsubsection{B[e] star candidates}

As discussed in Sect. 4.2 we have identified  a subset of objects with  IR excesses consistent with the hot dust that characterises sgB[e] 
stars and which are amongst the optically faintest of our sample (see  Thompson et al. \cite{thompson} for individual SEDs). 
While  this in part may be the result of a high degree 
of (circumstellar?) reddening towards certain objects, both  J013500.30+304150.9 and J013324.62+302328.4  appear to suffer little 
extinction  and hence are likely to be intrinsically faint (Fig. 5).   
Given the absence of photospheric features, determining bolometric corrections for both these stars
is difficult.  However, adopting values from Clark et al. (\cite{clark05b}) appropriate for the range 
of spectral types suggested for Magellanic Cloud sgB[e] stars  (B0 to B8; Zickgraf et al. \cite{zickgraf86}) we find, 
{\em prior} to applying an uncertain reddening correction\footnote{Ma07 give an average $E(B-V)=0.12$ for blue stars within M33 
which corresponds to $\sim$0.15dex in log$(L_{\ast})$; dwarfed by the uncertainty in the temperature dependant bolometric 
correction.},  log$(L_{\ast}/L_{\odot})\sim$4.1-5.1 and $\sim$4.2-5.2 for, respectively, J013324.62+302328.4 and
 J013500.30+304150.9. Such luminosities are  below  those of the sgB[e] stars analysed by Zickgraf et al. (\cite{zickgraf86})
and known Galactic LBVs (Clark et al. \cite{clark05a}), but  they are consistent with a lower luminosity population of stars demonstrating the B[e] phenomenon
 identified by Gummersbach et al. (\cite{gummersbach}).

What might be the physical nature of the B[e] star candidates? As described by Lamers et al. (\cite{lamers}), the B[e] phenomenon occurs in both single and
 binary stars  with a 
range of masses and evolutionary states. The lack of apparent molecular bands in the spectra of these stars would appear to exclude symbiotic systems, while
the mid-IR colours appear inconsistent with those of (massive) young stellar objects (e.g.  Bolatto et al. \cite{bolatto}). 
If a subset of stars are confirmed to be of low  luminosity (log$(L_{\ast}/L_{\odot}) \leq 4.7$), they would have corresponding 
 progenitor  masses of $\leq 15M_{\odot}$ (under the assumption they have left the main sequence; Meynet \& Maeder 
\cite{meynet}). While they may not be directly identified with the extreme
AGB stars identified by Thompson et al. (\cite{thompson}) - since their emission spectra imply they must be substantially hotter - such luminosities are comparable to those of the dust enshrouded 
progenitors of SN 2008S and the 2008 transient in NGC300, which are found to be optically obscured, mid-IR bright and non 
variable (Prieto \cite{prieto}, Prieto et al. \cite{prietoetal}, Thompson et al. \cite{thompson}). Consequently 
it is tempting to  consider the possibility that the 2 classes of object may be related, but subject to differing intrinsic 
absorptions; a suggestion first advanced for low luminosity  sgB[e] stars in the LMC  by Bonanos et al. (\cite{bonanos}). 

More luminous candidates would be comparable to the {\em bona fide} sgB[e] stars in the Magellanic Clouds. While  
the physical mechanism that leads to the production of sgB[e] stars is currently uncertain,  by analogy to classical Be stars it is suspected that
high rotation rates are necessary for the formation of a circumstellar disc. Both  binary interactions and mergers have been invoked to spin up the star,
 while a variant of this scenario - the presence of a circumbinary disc produced by short lived binary drive mass transfer - has been 
invoked to
 explain the sgB[e] star Wd1-9 (Ritchie et al. in prep). Given this, it is particularly interesting that radial velocity shifts led Sholukhova et al. 
(\cite{sholukhova04}) to conclude that  J013406.63+304147.8 is a short period binary, with mass loss in the equatorial plane leading to a sgB[e] classification.

\subsection{Individual objects} 

\subsubsection{Romano's Star}

\begin{figure}
\label{fig-fitroma}
\includegraphics[width=7cm,angle=90]{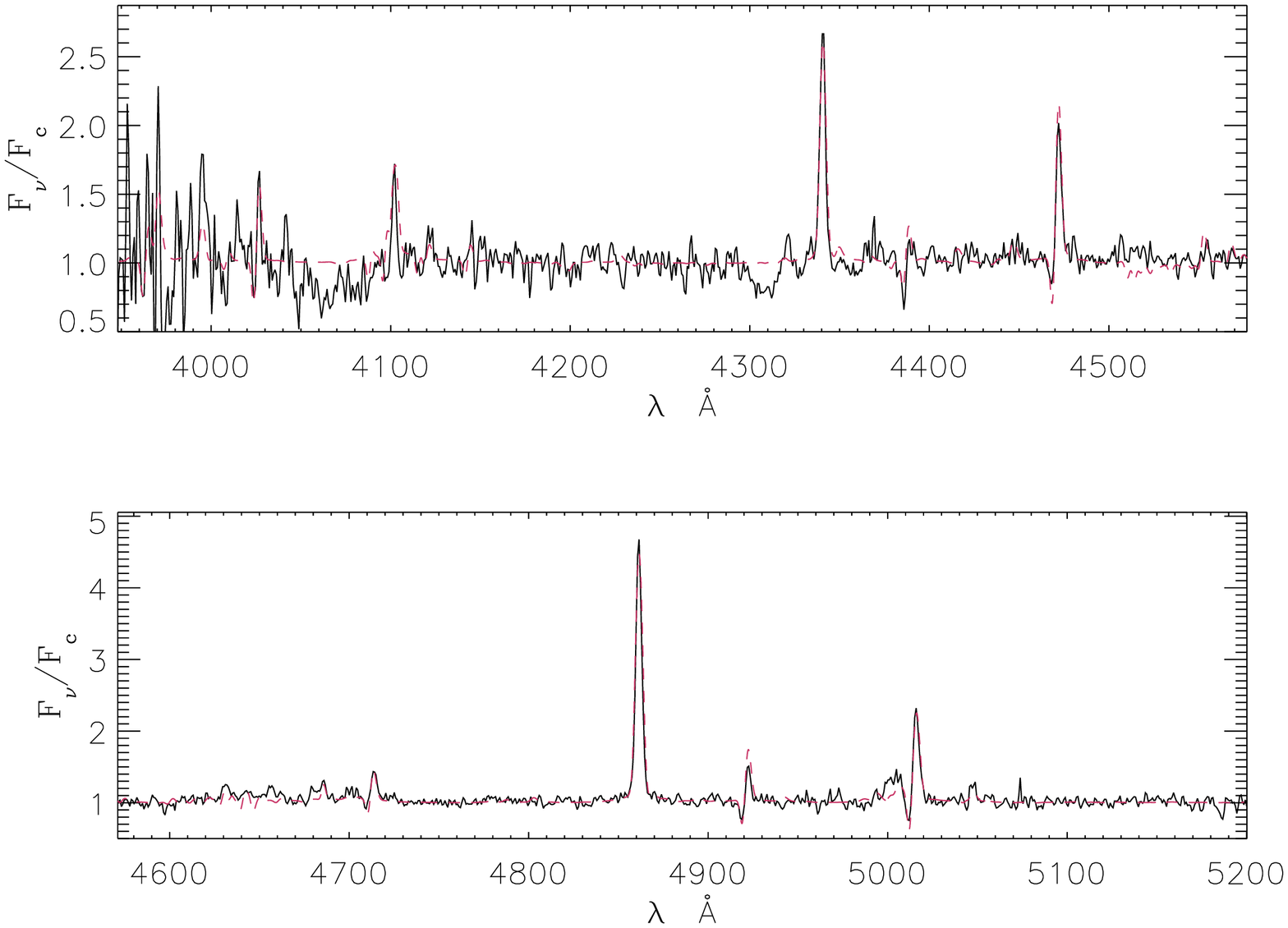}
\includegraphics[width=7cm,angle=90]{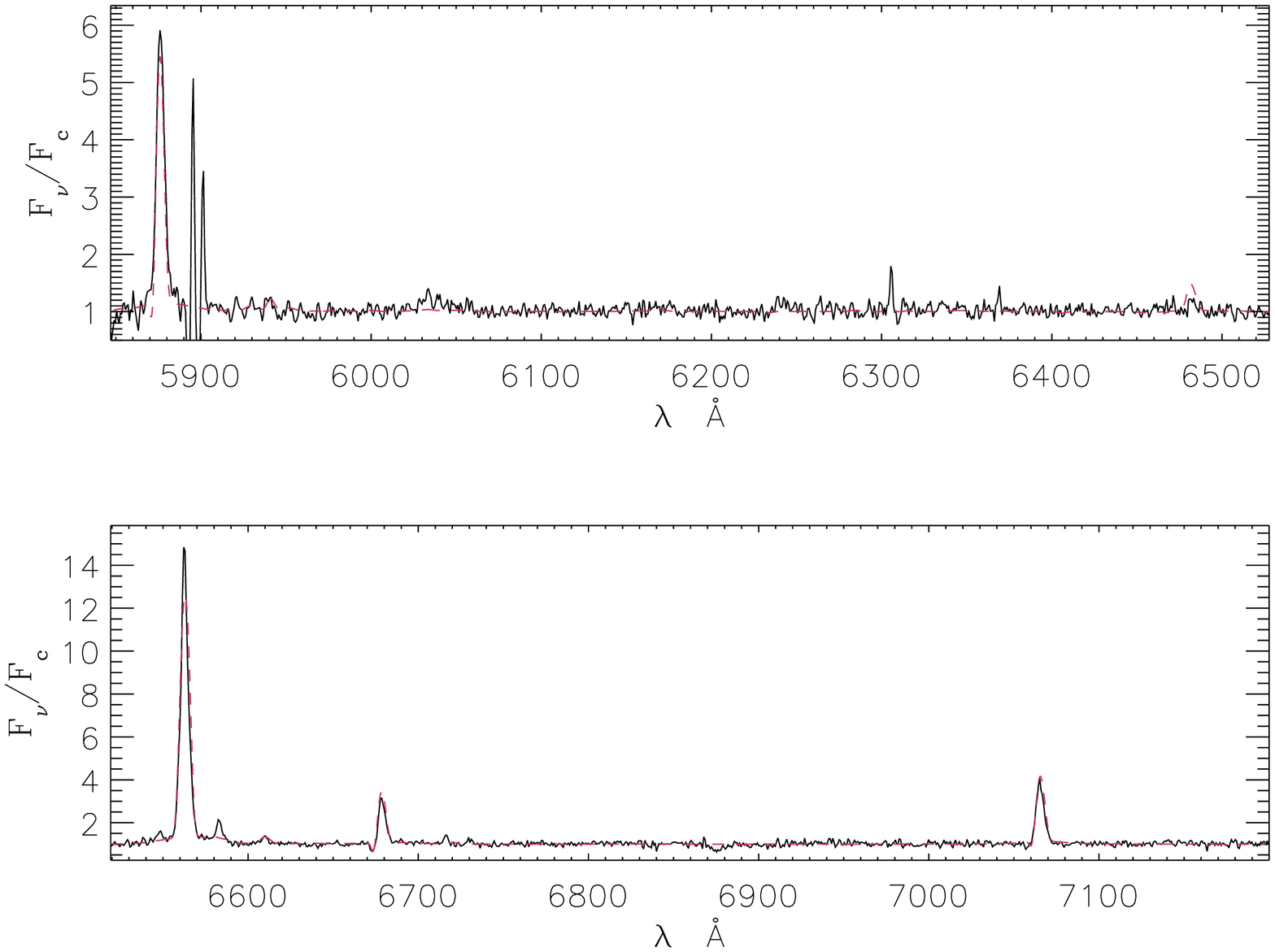}
\caption{Comparison of the synthetic spectrum (dashed) to the 2010 Sept. observations (solid) of Romano's star.}
\end{figure}

As highlighted in Sect. 3.4, observational studies clearly reveal the  LBV nature of Romano's star. Recent quantitative 
modeling of the star during the  optical minimum of 2007/8 and the local optical maximum in 2005 by Maryeva \& Abolmasov 
(\cite{maryeva}) reveal significant differences between the two epochs in the sense that the star was both hotter and more 
compact during optical minimum, as expected for canonical LBV excursions. Interestingly, however, these authors report a 
significant differnce in bolometric luminosity between the two phases, whereby the star was more luminous during 
optical {\em minimum}; a point we return to later. 

Given that a central goal of our program is to quantitatively follow the evolution of the stellar parameters of LBVs 
through their excursions/outbursts, Romano's star provided an ideal proof-of-concept; moreso since our spectroscopy and quasi 
simultaneous photometry (Sholukhova et al. \cite{sholukhova}) indicates that it appeared to be in a state intermediate 
between those described by  Maryeva \& Abolmasov (\cite{maryev11}). To accomplish this we utilised the non-LTE model 
atmosphere code CMFGEN (Hillier \& Miller \cite{h98}, \cite{h99}) and followed the methodology previously employed for the 
spectroscopically similar stars P Cygni (Najarro \cite{paco01}) and  
NGC~300 B16 (Bresolin et al. \cite{bresolin}), with the observational constraints
provided by our September 2010 blue and red spectra and the photometric values from Sholukhova et al. (\cite{sholukhova}). 

The latter were obtained  within one month of our spectrospic observations and according to Sholukhova et al.
 (\cite{sholukhova}), the object's $V$ magnitude varied between 17.75 and 17.85,
with $B$ ranging from 17.65 and 17.75, resulting in an average  $(B-V) \sim -0.10$ (Fig 2.b of Sholukhova et al. 
\cite{sholukhova}). During this epoch, the object was moving from the hotter, 
minimum brightness to the cooler, maximum brightness phase, being closer to the former. Fortunately, though rather weak, 
the \HeII$\lambda$4686 line could be used to constrain the ionization equilibrium and therefore derive the effective 
temperature of the star\footnote{We highlight that over the wavelength range discussed and  at the S/N of our observations, the He\,{\sc 
ii} 4686{\AA} line is
 the {\em sole} viable  
temperature diagnostic that permits spectroscopically similar objects such as P Cygni ($T_{\rm eff} \sim 18$kK; Najarro 
\cite{paco01}),  J013416.07+303642.1 and Romano's star ($T_{\rm eff}\sim 26$kK; Figs. A.3 and A.4) to be distinguished.
Spectra intended for quantitative  modeling of such objects  {\em must} therefore be of sufficient resolution and S/N 
to resolve this feature, while higher S/N observations would also permit N\,{\sc ii}/N\,{\sc iii} and Si\,{\sc iii}/Si\,{\sc iv}
ionisation equilibria to be employed.}.

Nevertheless, when combined with the ionization status displayed by the star, the moderate S/N of our spectra
 hamper an accurate determination of {\em all} stellar properties. Reliable estimates could be obtained for 
the effective temperature, \Teff, stellar luminosity, \Lstar, the wind terminal velocity 
\vinf, the H/He ratio and the mass-loss rate, \Mdot. On the other hand,  stellar parameters such as the exponent of 
the velocity field ($\beta$), degree of wind clumping ($f_{\rm V}$) 
and nitrogen and silicon abundances are subject to larger uncertainties, while only
 upper limits may be returned for additional  elements (e.g. Fe, C, Mg). In order to investigate this behaviour we
ran a suite of  models at solar, $0.7{\times}$ solar and $0.4{\times}$ solar metallicity and, apart from the Si lines, 
found no significant differences in either the quality of the final fit or the basic stellar properties,
noting  that higher S/N observations at maximum brightness (cooler phase) would unambiguously constrain the 
metallicity of the object. Because of this result  metallicity had to be assumed for our analysis. To accomplish this we utilized [O/H] 
estimates from the literature 
based on the radial position of the star within M33. According to Romano (\cite{romano}) the star is placed at a projected 
distance of 17' (4.8kpc) from the center of the galaxy, for which Neugent \&\ Massey (\cite{neugent}) and 
Bresolin (\cite{bresolin}) derive [O/H]${\sim}8.25-8.30$ and [O/H]${\sim}8.30$ respectively, corresponding to $0.4{\times}$solar. 
We adopted this value for the abundances of Na, Mg, Al, S, Ca, Ti, Cr, Mn, Fe, Co and Ni and assumed a distance of 
964kpc to M33 (Sect. 4.2)\footnote{Maryeva \& Abolmasov (\cite{maryev11}) adopt a slightly smaller distance (847~kpc) implying $\mu$=24.64. 
Correcting the stellar parameters for this lower distance  does not materially affect our conclusions.}.

The results of this analysis are summarised in  Table~\ref{tab:model}, while we compare our best-fit synthetic spectrum to 
observations in Fig. 8. As the emission strength of  
\HeII$\lambda$4686  is extremely sensitive to the effective temperature in this parameter domain we derive an
uncertainty of ${\sim}500$K for the \Teff\ of the object. Taking into account the uncertainties in the distance, photometry 
and reddening we estimate an error of 0.15dex for the stellar luminosity.
Wind properties are fairly well constrained. The terminal velocity, \vinf$=265\pm50$\kms, is set by the P Cygni 
profiles of the \HeI\ lines which, together with
the shape of the emission components of H$\alpha$ and H$\beta$, help constrain $\beta$. We adopted 
a two $\beta$-law with a flat $\beta=6$ component in the inner parts and a steeper $\beta=0.9$
law in the outer wind. Despite the medium S/N of our spectra, the strong electron scattering wings 
of H$\alpha$ point towards the presence of moderate clumping, $f_{\rm V}$=0.25. 

 Given the moderate spectral resolution of our 
observations we were not able to unambiguously identify and subtract potential nebular contamination of the H\,{\sc i} lines 
introduced by the fibre size (1.6" diameter; Sect. 2). To compensate for 
this we placed larger weight on reproducing the He\,{\sc i} and higher H\,{\sc i} Balmer lines than on H$\alpha$ and H$\beta$ when 
determining \.{M}. Thus our current favoured model slightly underestimates emission in the latter transitions and we estimate a 
likely  error of $\leq$25\% on this parameter in light of this uncertainty.
We obtain H/He=1.5 by number with  H/He=2.0 and H/He=1.2 as upper and lower limits respectively. 
The silicon abundance, with an uncertainty of 0.2dex, is basically constrained by the strength of the \SiIII$\lambda 4550-4575$ 
lines. Our derived nitrogen abundance reproduces satisfactorily the main optical \NII\ and \NIII\ lines but overestimates
significantly the \NII$\lambda$6482 line, which leads to a larger uncertainty (0.3dex) on the resulting abundance 
(a mass fraction X$_N=7\times10^{-3}$). Note that the [N\,{\sc ii}] nebular lines were not employed in this determination.
Upper limits to carbon and oxygen abundances of 0.07$\times$solar and 0.20$\times$solar were 
found, indicating a depletion in both elements.

Our results agree very well with those found by
Bresolin et al. (\cite{bresolin}) for the WNLh star B16 in NGC~300 and  encouragingly - and  as might be anticipated - 
land between the stellar parameters derived by 
Maryeva \&\ Abolmasov (\cite{maryev11}) for the minimum and maximum brightness
phases (thus  giving us confidence in our treatment of potential nebular contamination).
Our derived $E(B-V)=0.06$ is also consistent with the results of their analysis and yields a deredenned $V=17.62$.
 Interestingly, our derived Si abundance, which is roughly a factor of three lower than their value, agrees very 
well with the abundance of the other $\alpha$-elements  obtained from nebular
abundances at that galactocentric distance (Bresolin \cite{bresoli11}). The nitrogen enrichment and the upper limits on the C and O 
abundances are consistent 
with an WNL evolutionary phase, characterized by an LBV-like behaviour.

Multi-epoch,  quantitative analyses of the LBVs  AG Car (Groh et al. \cite{groh}), HD~5980 (Georgiev et al. \cite{georgiev}) 
and S Dor (Lamers \cite{lamersdor}) have been made during their  characteristic photometric excursions, and 
it is interesting to  compare the properties of Romano's star to these objects. As with AG Car and HD~5980, the mass loss 
rate is found to increase and wind velocity decrease with decreasing temperature (Lamers  \cite{lamersdor} 
does not report these parameters for S Dor). However, the bolometric luminosities of AG Car, HD~5980 and S Dor are 
found to {\em decrease} with the simultaneous reduction in
temperature and increase in stellar radius, whereas the opposite appears to be the case for Romano's star (Mareya \& Abolmasov
\cite{maryev11}).

It is thought that the reduction in bolometric luminosity for the first  three objects results 
from  the additional energy required to support the expansion of the outer layers of the stars; if correct
 this would 
imply that the overal energy budget of Romano's  star {\em increased} during its excursion. In this regard we highlight that 
the behaviour of the Galactic  LBV AFGL2298 also implies an increase in the rate of energy production during its `eruption', although in this case 
the growth in stellar radius occured at $\sim${\em constant} temperature (Clark et al. \cite{clark09}). If the results of these
studies are replicated for these and other stars, the clean demarcation between luminosity conserving excursions and 
non-luminosity conserving eruptions in LBVs will break down, despite the superficial similarities in their 
spectral and photometric evolution (cf. AG Car, S Dor and Romano's star); consequently raising the question as to 
whether a single physical mechanism underlies their divergent behaviour.

\begin{table*}[t]
\caption{Stellar parameters derived for Romano's star.}
\label{tab:model}
\begin{center}
\begin{tabular}{cccccccccc} 
\hline\hline
   $M_V$    & log(\Lstar)  & $R_{2/3}$  & \Teff  & \Mdot              & \vinf  &  \fv  & H/He & N/N$_{\odot}$ & Si/Si$_{\odot}$\\
          & $L_{\odot}$       &  \Rsun     &  kK    &  $10^{-5} M_{\odot}$yr$^{-1}$ & \kms   &       &      &               &               
\\
\hline
  $-7.27$  &  5.85        &  41.5     &  26.0  &  2.18               &  265   &  0.25 &  1.5  &  6.4         &  0.50         \\
\hline
\end{tabular}
\normalsize
\end{center}
{Note that the H/He ratio is given by number. Abundances relative to solar values assume solar values
from Anders \&\ Grevesse (\cite{anders89})}
\end{table*}

\subsubsection{M33 Var C}

M33 Var C has been the subject of considerable photometric and spectroscopic  attention since its discovery (Hubble  \& Sandage \cite{hubble}), albeit of
irregular temporal coverage.  Burggraf et al. (\cite{burggraf}) present the most complete long term ($\sim$1900-2010)
B-band optical light curve available, with observations reported in Humphreys et al. (\cite{humphreys88}) and Szeifert et al. (\cite{szeifert}) extending the 
wavelength coverage to the $V-$ and $R-$bands. These data reveal two prominent photometric maxima ($B \sim 14-15$, $V \sim 15.2$) between 1945-50 and 1984-8. Outside 
of these events, the star was typically observed with  $B \sim 16-17.5$ ($V \sim 16-17$)  although for at least 2 decades prior to its initial brightening in  
1940 it appeared to be in an extended faint phase ($B \sim 17.5-18$, $V \geq17$). As can be inferred from the magnitude ranges quoted, significant short term 
variability also appears to be present (e.g. Fig. 4)  in addition  to the  uncertainties associated with the 
historical observations. Nevertheless M33 Var C currently appears to be in an extended  bright phase which commenced in $\sim$1940 and upon which are superimposed two 
further outbursts separated by $\sim$40~yr (Burggraf et al. \cite{burggraf}). The somewhat sparse post-1990 photometric data presented by these authors appear to 
indicate a plateau state ($B \sim 16.4$) from $\sim$1990-2005 after which M33 Var C faded by $\sim$0.8mag. 

Spectroscopic observations are likewise uneven. During both  visual  maxima M33 Var C demonstrated
  an F-supergiant  plus Balmer emission line spectrum analagous
to S Dor in its bright phase    (Hubble \& Sandage \cite{hubble}, Humphreys et al. \cite{humphreys88}). Outside of these events,  the limited spectroscopy appears 
to indicate an earlier B spectral type dominated by Fe\,{\sc ii} and Balmer line emission - cf. the most recent high S/N and resolution spectrum of Szeifert 
et al. (\cite{szeifert}), dating from 1992. More recently, Viotti et al. (\cite{viotti}) suggest a similar appearance in 2002 while, somewhat 
surprisingly, Burggraf et al. (\cite{burggraf}) claim - but do not show - a much later spectral type coinciding with the current photometric fade.

The spectroscopic and photometric data presented here suggest a hitherto unreported excursion analagous to the 1940s and 1980s events. Commencing 
in $\sim$2001 July  the optical peak  was reached a year later and was maintained for a further year,  during which M33 Var C demonstrated the 
F-supergiant emission  line spectrum that characterised previous outbursts. If our interpretation of the mid-IR emission is correct (Sect. 4.2) this maximum light 
plateau endured until 2005 February. The limited sampling of the B-band lightcurve of Burggraf  et al. (\cite{burggraf}) meant that they completely missed this event; unfortunately
 it casts doubt on their 
conclusion that the bright phases of M33 Var C occur on a $\sim$40~yr period.
Moreover, these data appear to demonstrate that  spectral variability in M33 Var C may occur over relatively short timescales, with an F-hypergiant spectrum 
observed in 2003 (this 
work), B emision line in 2004 (Viotti et al. \cite{viotti}), late F in $\sim$2008 (Burggraf et al. \cite{burggraf}) and B emission in 2010. For comparison recent 
complete outburst cycles (hot to cool phase and back again) of AG Car and HR Car occured over $\sim$10 and  $\sim$8~years respectively (Stahl et al. 
\cite{stahl01},
Szeifert et al. \cite{szeifert03}).

\subsubsection{[HS80] B324}

Visually one of the brightest stars within  M33, B324 is discussed at length by Monteverde et al. 
(\cite{monteverde}) who highlight both its late, possibly variable  spectral type  (F0-F5 Ia$^+$ in  1994 versus A5 Iae prior 
to 1990; Humphreys et al. \cite{humphreys90}) and LPV in the Balmer series. P Cygni
lines are observed in the FeII multiplets 40 and 70, which require a dense and ionised medium to be driven into 
emission. Such a morphology is rare, being demonstrated by both M33 Var B and C (Szeifert et al. \cite{szeifert} and this work) during their cool 
photometric maxima and the Galactic yellow hypergiant (YHG) IRC+10 420 (Fig. 9), a star  thought to be transiting rapidly to hotter temperatures
as it executes a red loop across the HR diagram.

However, the similarities to IRC+10 420 appear to end here. Prior to the 1990s IRC +10 420 underwent a long term photometric brightening (Patel et al. \cite{patel})
and while B324 also appears photometrically variable (eg. Ma07) the limited data are consistent with $\alpha$ Cygni variability.
Moreover IRC +10 420 is associated with significant dusty circumstellar ejecta of recent origin leading to a substantial IR excess (Bl\"{o}cker et al. \cite{blocker}); in contrast, despite its intrinsic luminosity the mid-IR colours of B324 do not appear consistent with other 
massive stars experiencing, or having recently undergone, heavy mass loss (Fig. 6)\footnote{We suspect the IR emission of B324 arises from a combination of  photosphere+wind, although 
interpreting the properties of
such cool stars is not possible utilising current model atmosphere codes such as CMFGEN. For example, while we were able to reproduce the photospheric absorption 
spectrum of the LBV Wd1-243 in the cool phase we were unable to simultaneously reproduce the emission lines originating in the stellar 
wind (Ritchie et al. \cite{w243}).}.
Finally, B324 appears significantly more luminous than IRC +10 420. Adopting  $V=14.86$ and $(B-V)=0.43$ (Massey et al. \cite{massey06}) and 
the intrinsic colours of a F5 Ia star ($(B-V)_{\rm o}=-0.15$; 
Fitzgerald \cite{fitzgerald}) leads to $E(B-V)=0.28$ and $A_V=0.87$ (assuming $A_V=3.1E(B-V)$ and subject to previous caveats regarding the 
uncertainty in dereddening individual stars). Then
 assuming a negligible bolometric correction and $\mu$=24.92 we find log$(L_{\ast}/L_{\odot}) \sim 6.3$, placing it significantly in excess of the 
empirical Humphreys-Davidson limit.

Such a luminosity appears to be incompatable with a star evolving through a YHG phase via a red loop across the HR diagram, suggesting instead that 
an identification with a cool phase LBV outburst may be more appropriate (cf. M33 Var B and C). A consequence of this is that since there is currently no 
observational evidence for long term 
secular variability, B324 appears to have remained in its current extended `outburst' phase for at least twenty years. However, decades long outbursts have been observed in both the cool hypergiant
M33 Var A ($\geq$35~yr; Humphreys et al. \cite{humphreys06}) and R127 ($\sim$30~yr ; Walborn et al. \cite{walborn08}), while M33 Var C 
has yet to return to its 
pre-1940 quiescent state, suggesting that such behaviour is not without precedent. Intriguingly, the BHG [HS80] 110A is also located above the HD limit 
(log$(L/L_{\odot}){\sim}6.5{\pm}0.1$ and $T_{\rm eff}\sim 21{\pm}1$kK; Urbaneja et al. \cite{urbaneja05}) and shows no evidence for LBV-like variability over the past $\geq 17$yr (Sect. 3.3). Likewise, 
the Galactic BHG Cyg OB2 \#12, which  is also found to  violate the HD limit, has remained stable for over a century (Clark et al. \cite{clark11b}).

\begin{figure}
\includegraphics[width=7cm,angle=270]{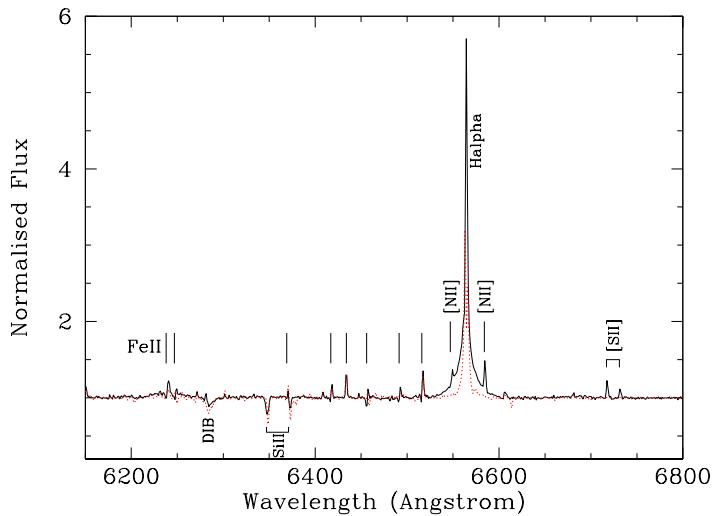}
\caption{Comparison of the 1996 spectrum of the emission line F hypergiant B324 (solid line; Monteverde et al. \cite{monteverde})
against the Galactic YHG IRC +10 420 degraded to a comparable spectral 
resolution (red dotted line; Oudmaijer et al. \cite{oudmaijer}). Similar Fe\,{\sc ii} P Cygni emission lines are also observed in the 2003 spectrum of M33 Var C.}
\end{figure}

\section{Global considerations and concluding remarks}

Finally, what may we infer from the current population of candidate LBVs within M33? Together with the photometric selection criteria employed by
Ma07, multiepoch observations employing  wide field imagers and multiobject spectrographs on 4-m class telescopes appear ideally suited to their 
study, as illustrated by the identification of a third LBV excursion of M33 Var C between 2001-5.
Combining existing spectroscopic and photometric datasets reveals a number of new  LBV {\em candidates} in addition to the  well studied
 historic examples (e.g. Szeifert et al. \cite{szeifert}, Viotti et al. \cite{viotti}, Ma07).  The dramatic 
spectroscopic variability of the P Cygni star  J013339.52+304540.5 and the WNL star J013332.64+304127.2 (Ma07) makes a compelling case for both objects.
Pronounced long term photometric variability is present in the BHG  J013429.64+303732.1 and the iron star J013459.47+303701.0. 

Moreover, low level optical photometric and spectroscopic variability appears ubiquitous amongst our sample, with LPV in the wind lines present
 in examples of all the (modified) classifications 
introduced by Ma07 and in the  photospheric lines of the BHGs (particularly the Mg\,{\sc ii}/He\,{\sc i} diagnostic lines). Unfortunately, 
our current temporal sampling is insufficent to distinguish between an origin of the  former behaviour in short lived stochastic or long lived LBV driven 
changes in the properties of the wind, while the latter could  arise in  photospheric  pulsations, secular evolution of stellar temperature and/or 
changes in the structure of the wind/photospheric boundary. Nevertheless, optical photometry 
strongly suggests that many of the stars are $\alpha$ Cygni variables and hence likely pulsationally unstable. 

We find that the spectroscopic data are also of sufficient quality to permit non-LTE model atmosphere analysis 
and we present a proof-of-concept study of Romano's star. The successful analysis of this star demonstrates the 
efficacy of our observational approach, but emphasises that high S/N and resolution data are mandatory to distinguish
 between Ofpe/WNL and P Cygni classifications and to determine the temperature of the BHGs (e.g. via the 
Si\,{\sc ii}:Si\,{\sc iii} ratio). Moreover, more extensive spectral observations would be required to ascertain the 
full elemental composition of such stars as well as to drive down the uncertainties on parameters such as the wind clumping  
factor. 

As expected from the spectroscopic data - being  intermediate between  those obtained in  photometric minima and maxima 
- we found that the physical  properties of Romano's star in 2010 
 imply a smooth progression between these extremes. We note that the  increase in bolometric luminosity with 
increasing radius and decreasing temperature  is in the {\em opposite} sense to that demonstrated by AG Car, 
S Dor  and HD~5980, in which  bolometric luminosity is observed to decrease. In combination with analysis of AFGL2298 this 
 suggests that the simple division  of LBV behaviour into bolometric luminosity conserving excursions
 and non-conserving eruptions is overly simplistic.

As suspected by Ma07, the subset of iron stars 
appear to be  rather heterogeneous containing, in addition to strong LBV candidates,
a  number of objects that can formally be classified as B[e] stars on the basis of their optical  and mid-IR properties. 
Their nature is currently uncertain,  but the lower luminosity subset may be related to the class of dusty, highly obscured objects identified 
by Thompson et al. (\cite{thompson}) as potential progenitors of transients such as SN 2008S, albeit suffering less 
self extinction. These authors find that 
approximately 25 of the 45 objects with similar mid-IR colours and magnitudes to the low luminosity B[e] stars 
(M$_{[4.5]}<-11.5$ and 0.5$<$[3.6]$-$[4.5]$<$1.5) are likewise 
optically faint or invisible, suggesting that they 
comprise an important component  of the population of the most IR luminous stellar objects within M33.

 Of the remaining iron stars,  N93351, N125093, J013350.12+304126.6 and J013442.14+303216.0
  possess optical spectra consistent with confirmed LBVs and mid-IR properties (colours and magnitudes) 
approaching  known massive evolved stars with extensive recent or ongoing mass loss; a connection  with the  recently discovered 
`Object X' - the brightest stellar mid-IR  source detected within M33 (Khan et al. \cite{khan}) - would be unsurprising. 
No stars appear to be  as mid-IR luminous as $\eta$ Carinae {\em currently} is, despite over a century elapsing
since  its eruption. While this might be considered evidence of the rarity of comparable events, such a conclusion is hampered by observational
 bias - one might expect highly luminous stars to be located in active star forming regions that may themselves be IR bright leading to a lack 
of detection due to source confusion. Five further stars have 8$\mu$m fluxes that are suggestive of the presence of dusty ejecta.
While the {\em bona fide} LBVs M33 Var C, Romano's star and spectroscopic candidates J013339.52+304540.5 and J013332.64+304127.2
 lack near-mid IR excesses  we 
 may not at present  exclude the presence of ejection nebulae associated with these stars,  since emission from such dusty structures
 surrounding  Galactic LBVs typically peaks at longer wavelengths.

Adopting the modified classification scheme of Ma07 and excluding the non-LBV subset of BHGs and iron stars we infer
a progression of Ofpe/WNL $\rightarrow$ P Cygni $\rightarrow$ iron star/BHG $\rightarrow$ cool hypergiant for the  transition from high 
to low temperatures. Due to the empirical luminosity dependance of the amplitude of LBV excursions, lower luminosity examples may 
not reach the Ofpe/WNL and P Cygni phases, while  some well studied LBVs such as AG Car have yet to be
 observed in a cool hypergiant phase (Stahl et al. \cite{stahl01}). Conversely, Romano's star and the  SMC 
system HD~5980 (Koenigsberger et al. \cite{koen} and refs. therein) suggest that  LBV-like excursions 
may  also extend to somewhat earlier WN subtypes than employed in this scheme ($\leq$WN8), while some cool hypergiants such as M33 Var A and $\rho$ Cas
are instead  observed to transit to cooler temperatures in outburst.

In conclusion, such galaxy wide multiepoch  observations employing large sample sizes appear well suited to the determination of  the 
duty cycle and physical nature  of  the outbursts  and  eruptions of LBVs and related objects, 
as well as   the length of the LBV phase (via comparison of number counts 
to main sequence, Wolf-Rayet and cool super-/hypergiant populations). However, several caveats must be considered. 
Firstly, any such survey should ideally extend to earlier and later spectral types than included in 
the classification scheme of Ma07 in order to sample the full range of excursions undertaken by luminous evolved stars.
Moreover, given the apparent LBV excursion of J013459.47+303701.0 (with $V\sim18.4$) any study must also to extend to comparatively faint  
candiates which are currently 
poorly sampled. 

 Combined with historical data, these observations support the finding that some excursions/outbursts of luminous evolved stars may last several decades. Both J013350.12+304126.6 and B324  have spectra that resemble those of LBVs in 
outburst but neither have shown 
evidence of secular variability over the last $\sim$20~years. Likewise M33 Var C   has yet to return to its pre-1940s quiescent state and 
consequently  
it {\em could} be argued to have been  in `outburst' for 
$\sim$70~yr. Indeed, it appears possible that current assumptions as to the typical timescale of LBV excursions and eruptions have been   biased to shorter 
durations by the length of the typical astrophysical  career (as well as a small sample size). If such an assertion is correct, a considerable long term 
observational investment will need to be made
in order to understand such behaviour, and its relation to the more rapid ($\leq$yr) `outbursts' and pulsations that are exhibited by cooler transitional 
objects such as YHGs and RSGs 
(e.g. Clark et al. \cite{clark10}).  Finally, despite their apparent stability,  both B324 and the BHG [HS80] 110A are found to exhibit combinations of temperature and luminosity which place them in 
excess of the HD-limit, joining a handful of other stars such as Cyg OB2 \#12 that also appear to violate it; intriguingly the latter object also appears to have 
remained stable for over a century (Clark et al. \cite{clark11b}).

\begin{acknowledgements}
This research is partially supported by the Spanish Ministerio de
Ciencia e Innovaci\'on (MICINN) under
grants AYA2008-06166-C03-02/03,  AYA2010-21697-C05-01/05  and Consolider-Ingenio 2010 project CSD2006-70 and the project of the 
Gobierno de Canrias PID2010119. We thank the referee for their careful reading of the manuscript which lead to significant improvement 
in its presentation.
We also thank Chris Evans for facillitating the collaboration
that led to Fabio Bresolin and Norbert  Pryzbilla kindly providing their unpublished Keck/DEIMOS 
spectra,  John Hillier for providing the CMFGEN code, Phil Massey for providing 
published spectra in electronic format, Jos\'{e} Prieto for providing unpublished 
mid-IR data and Guy Stringfellow for providing comments on the final version.  Finally, JSC would like to dedicate this paper to 
his new daughter, Freya, and her mother, Laura.

\end{acknowledgements}

{}

\appendix

\section{Presentation of spectroscopic data}

\begin{figure*}
\includegraphics[width=12cm,angle=270]{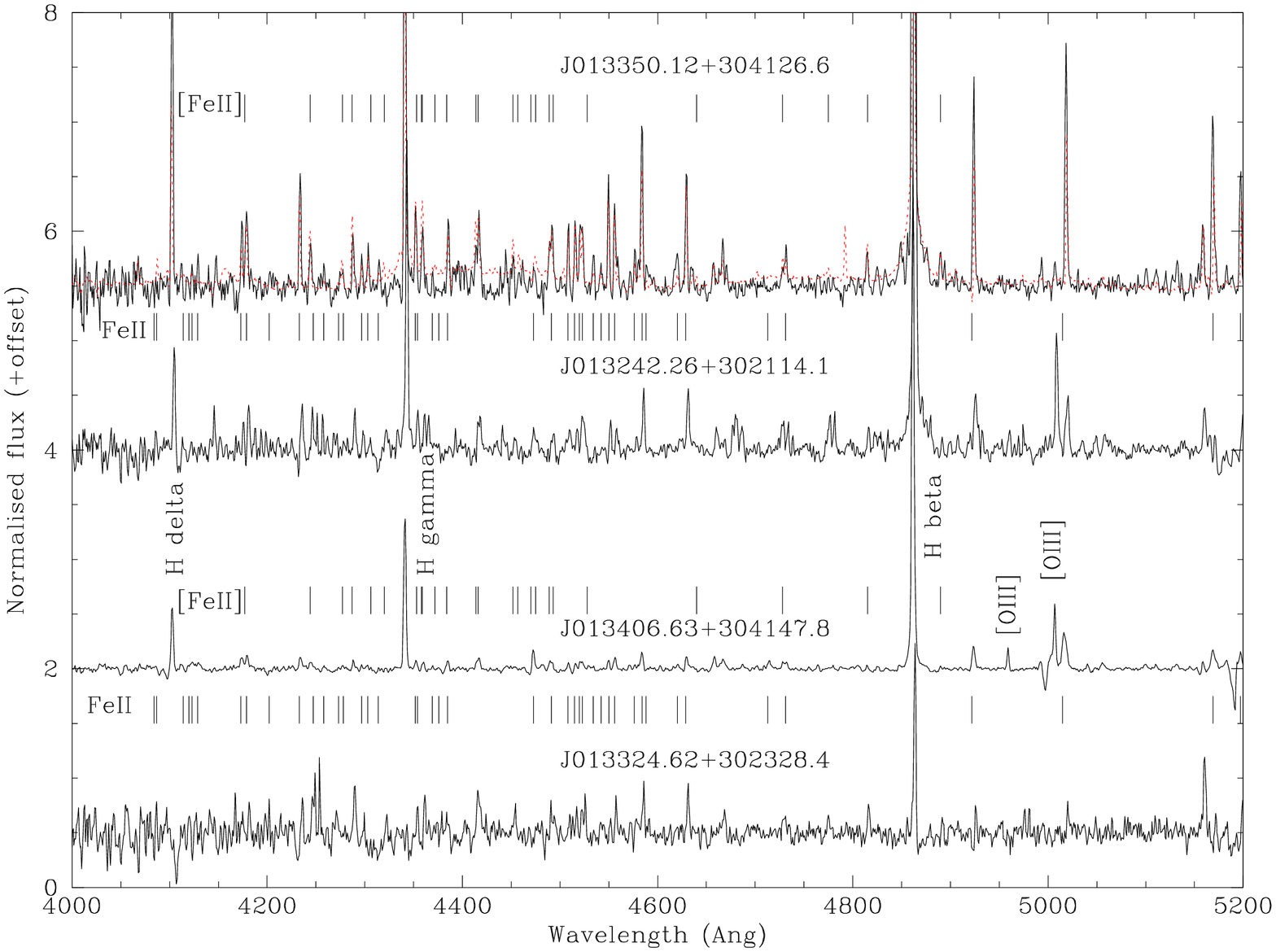}
\includegraphics[width=12cm, angle=270]{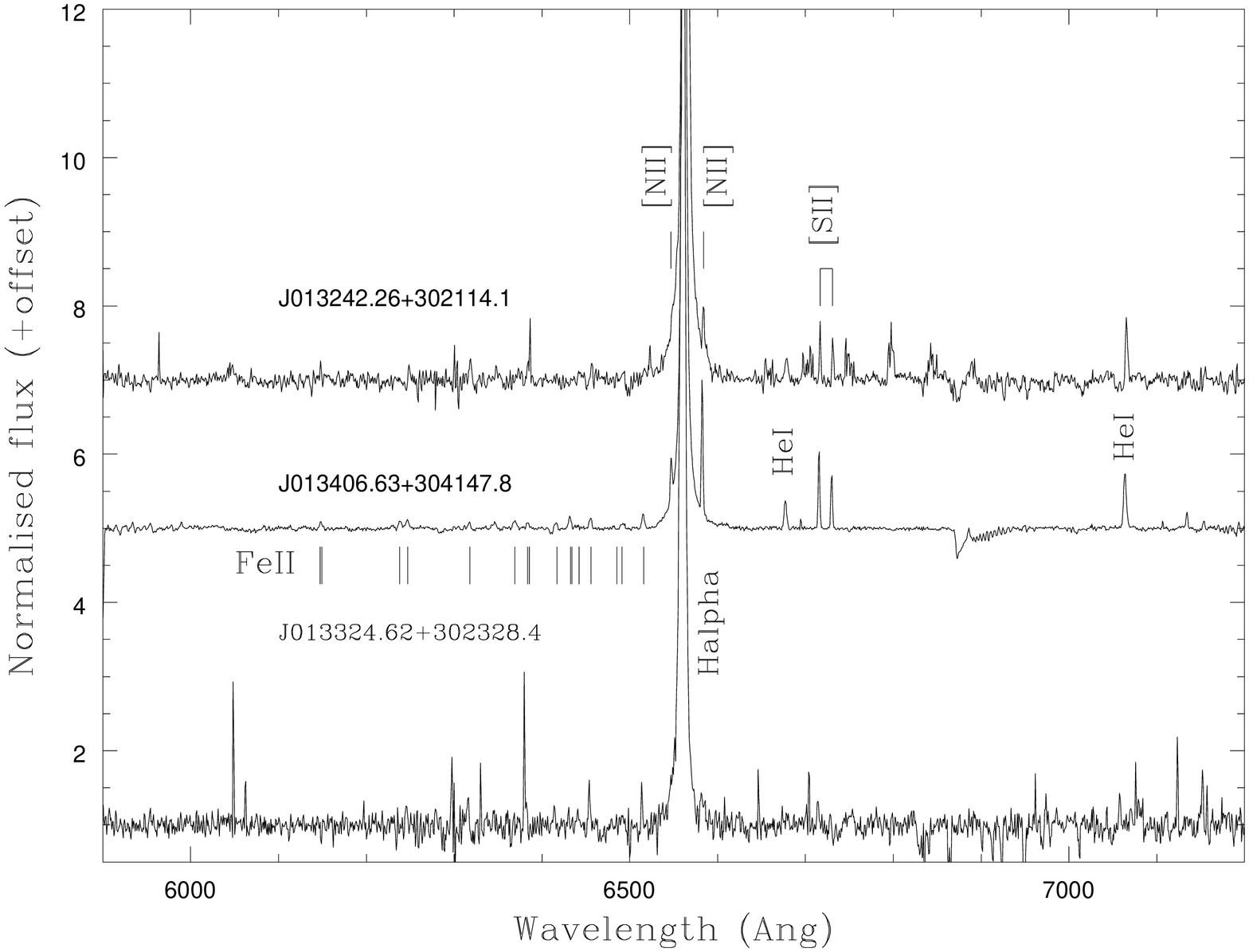}
\caption{Optical spectra of iron stars. A  spectrum of the LMC LBV R127 obtained
in during outburst in 1999 July is overplotted 
(red dashed line) on the spectrum of J013350.12+304126.6 for comparison,
 after having the 
resolution degraded to match.}
\end{figure*}

\begin{figure*}
\includegraphics[width=12cm,angle=270]{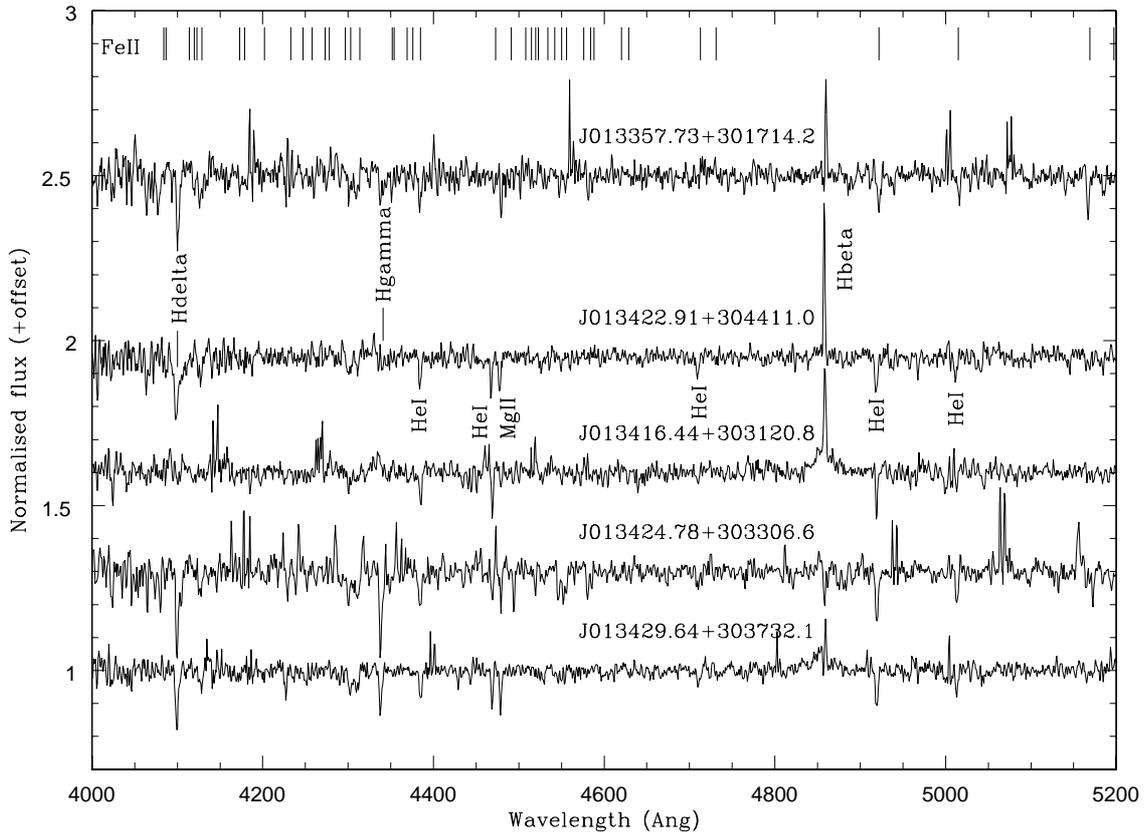}
\includegraphics[width=12cm,angle=270]{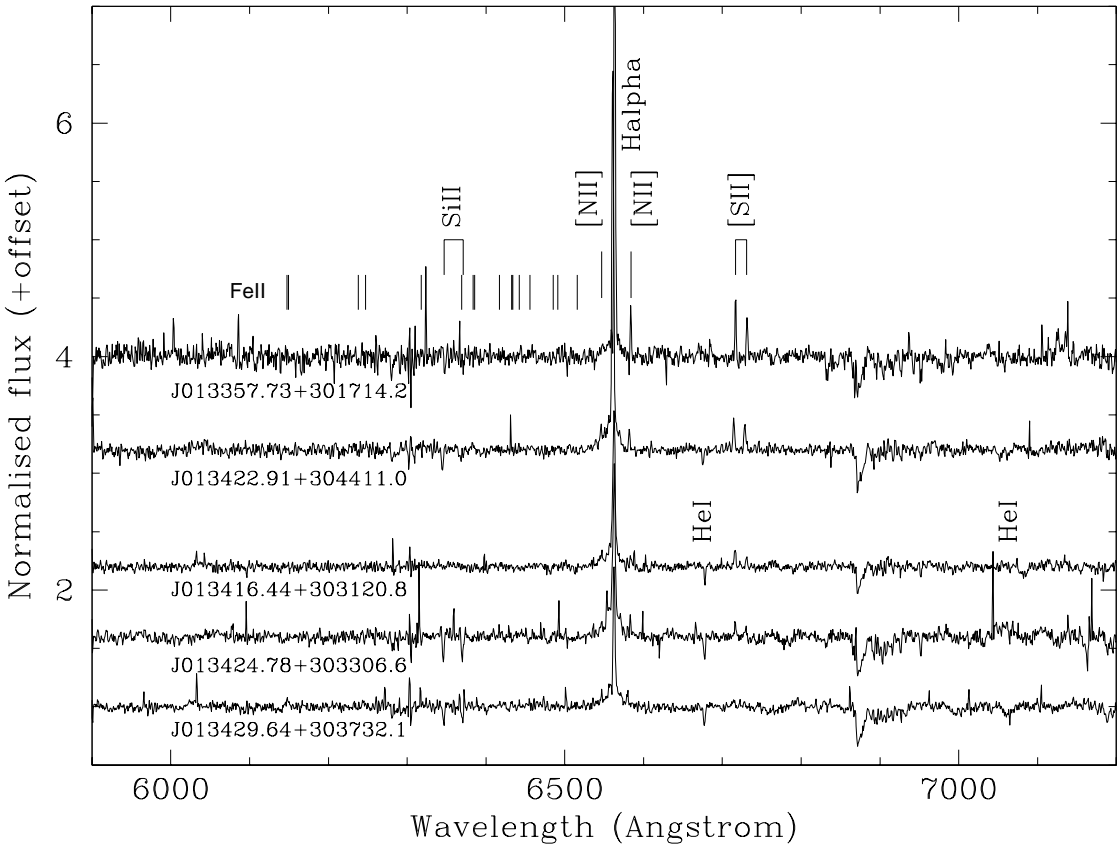}
\caption{Optical spectra of BHGs.}
\end{figure*}

\begin{figure*}
\includegraphics[width=12cm,angle=270]{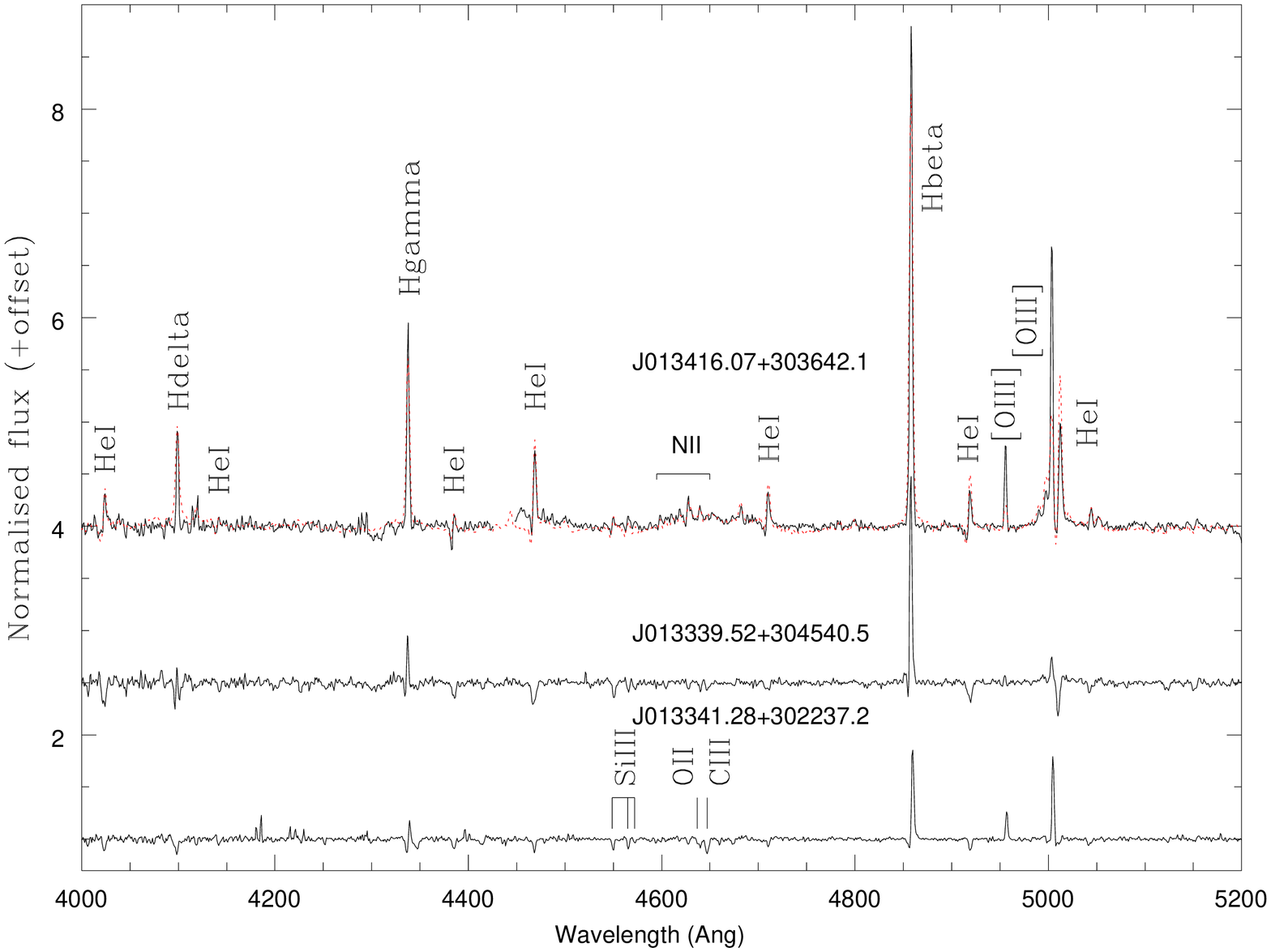}
\includegraphics[width=12cm,angle=270]{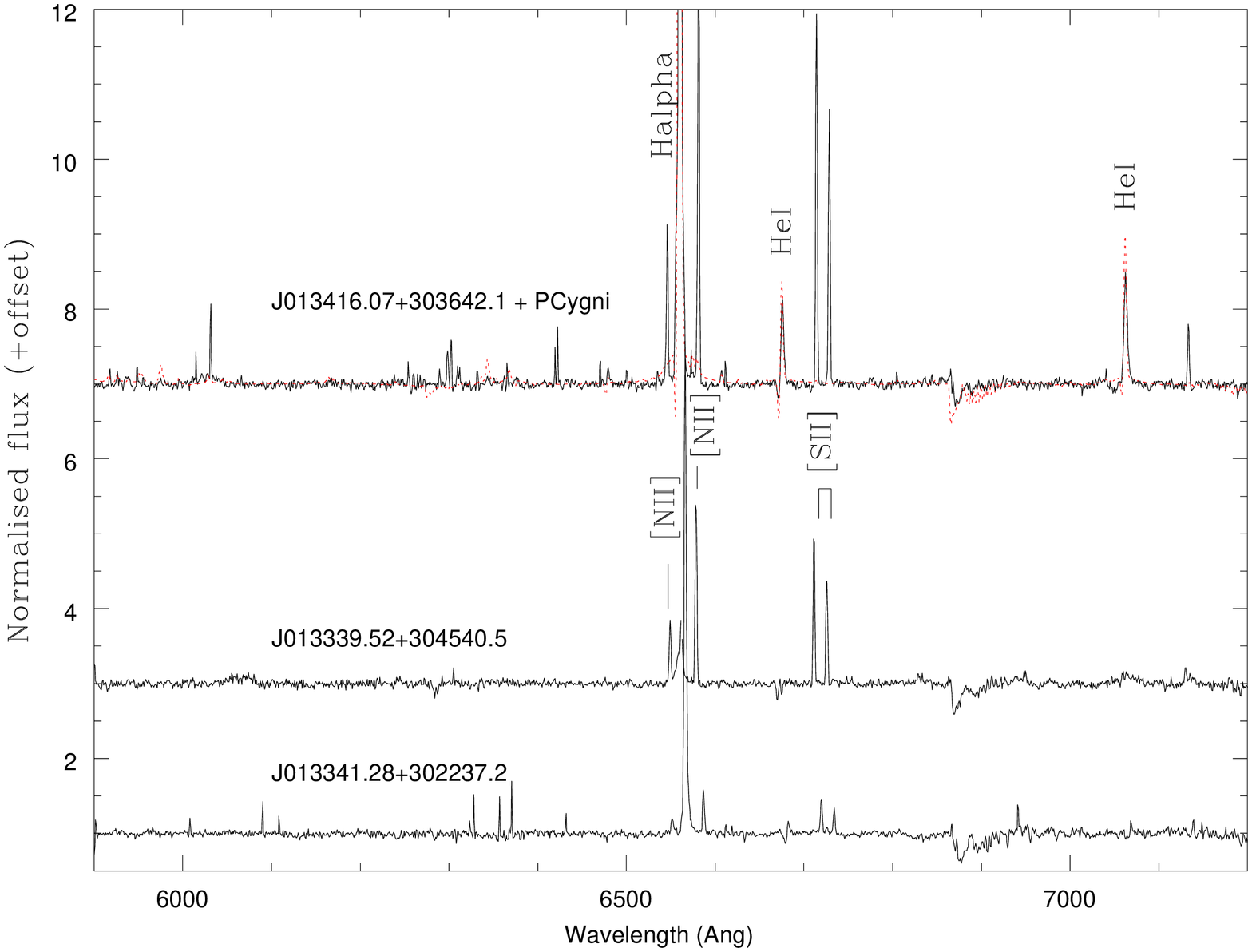}
\caption{Optical spectra of candidate P Cygni LBVs. A high resolution spectrum
 of P Cygni itself is   overplotted 
(red dashed line) on the spectrum of J013416.07+303642.1 after having the 
resolution degraded to match.}
\end{figure*}

\begin{figure*}
\includegraphics[width=12cm,angle=270]{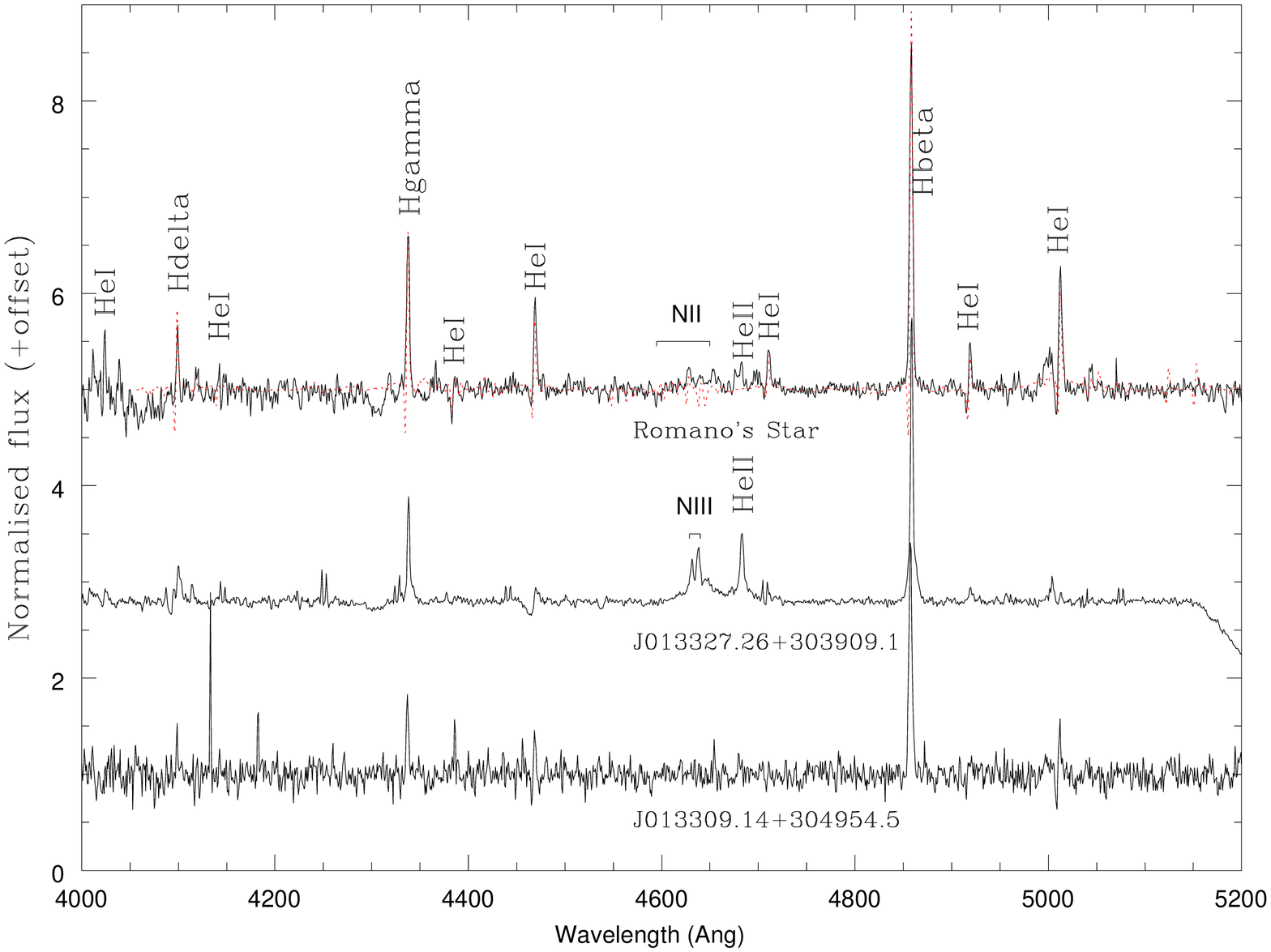}
\includegraphics[width=12cm,angle=270]{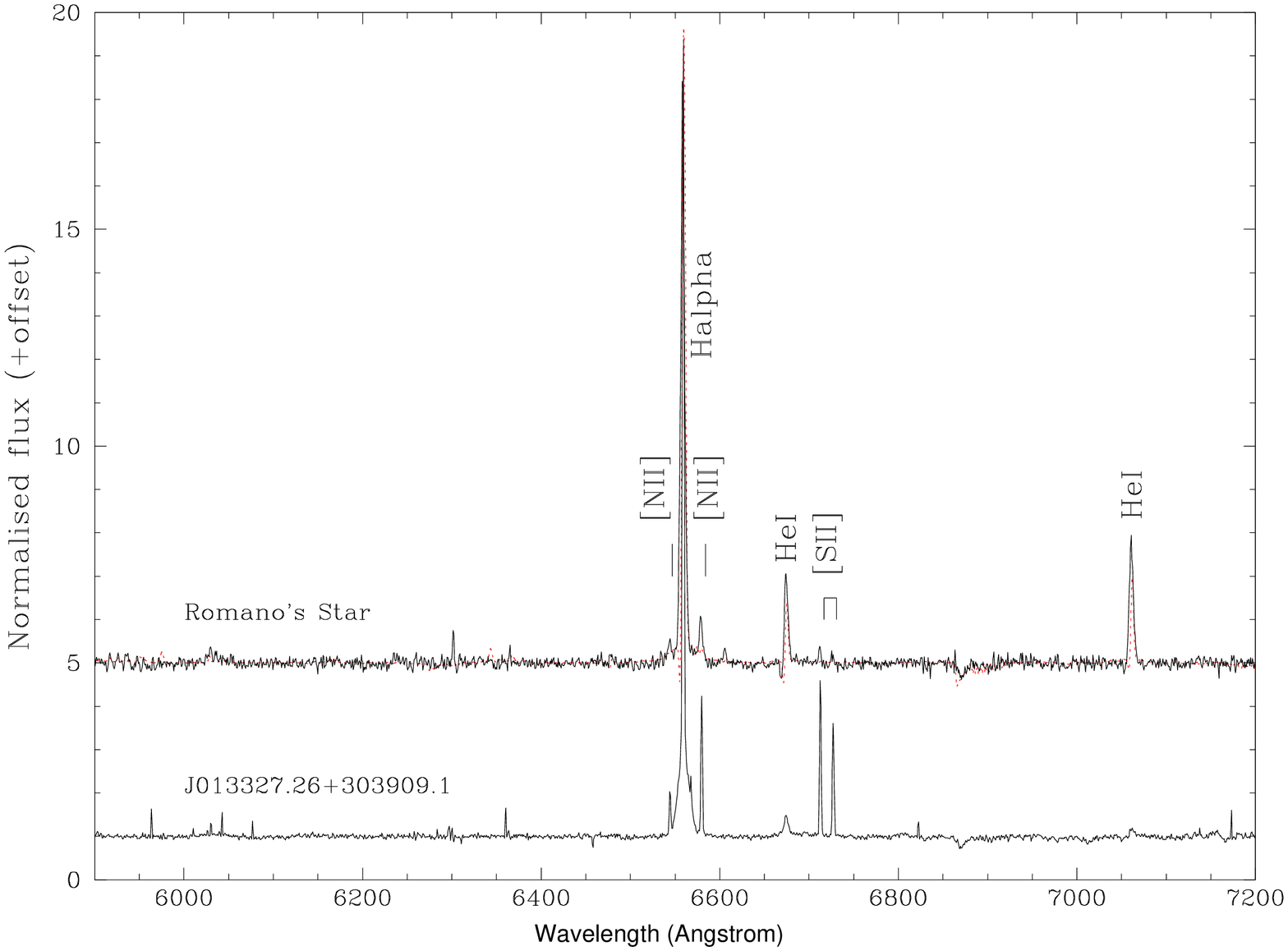}
\caption{Optical spectra of candidate Ofpe/WNVLs. 
A  spectrum of P Cygni is overplotted 
(red dashed line) on the spectrum of Romano's star for comparison,
 after having the 
resolution degraded to match.}
\end{figure*}

\begin{figure*}
\includegraphics[width=12cm,angle=270]{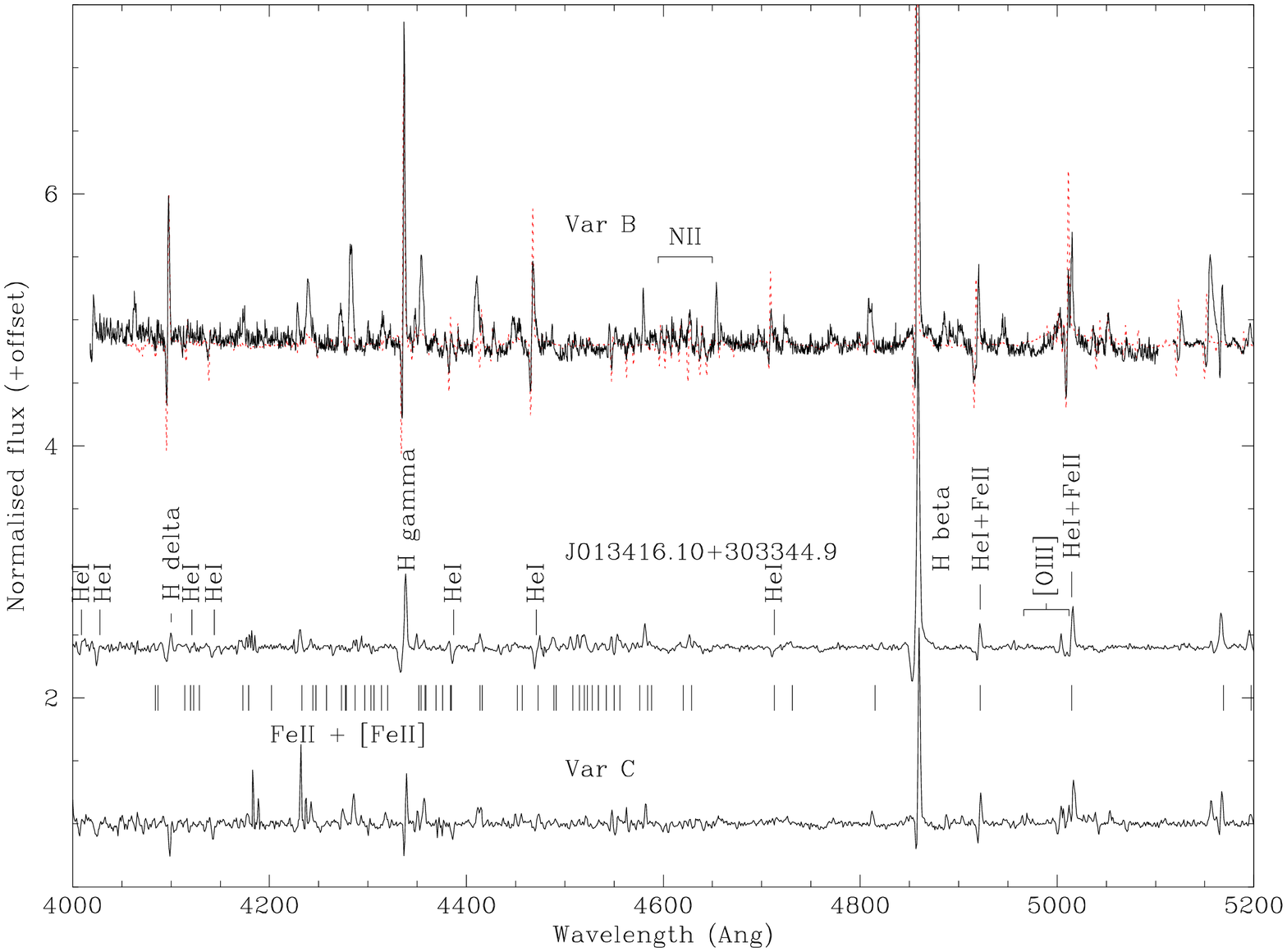}
\includegraphics[width=12cm,angle=270]{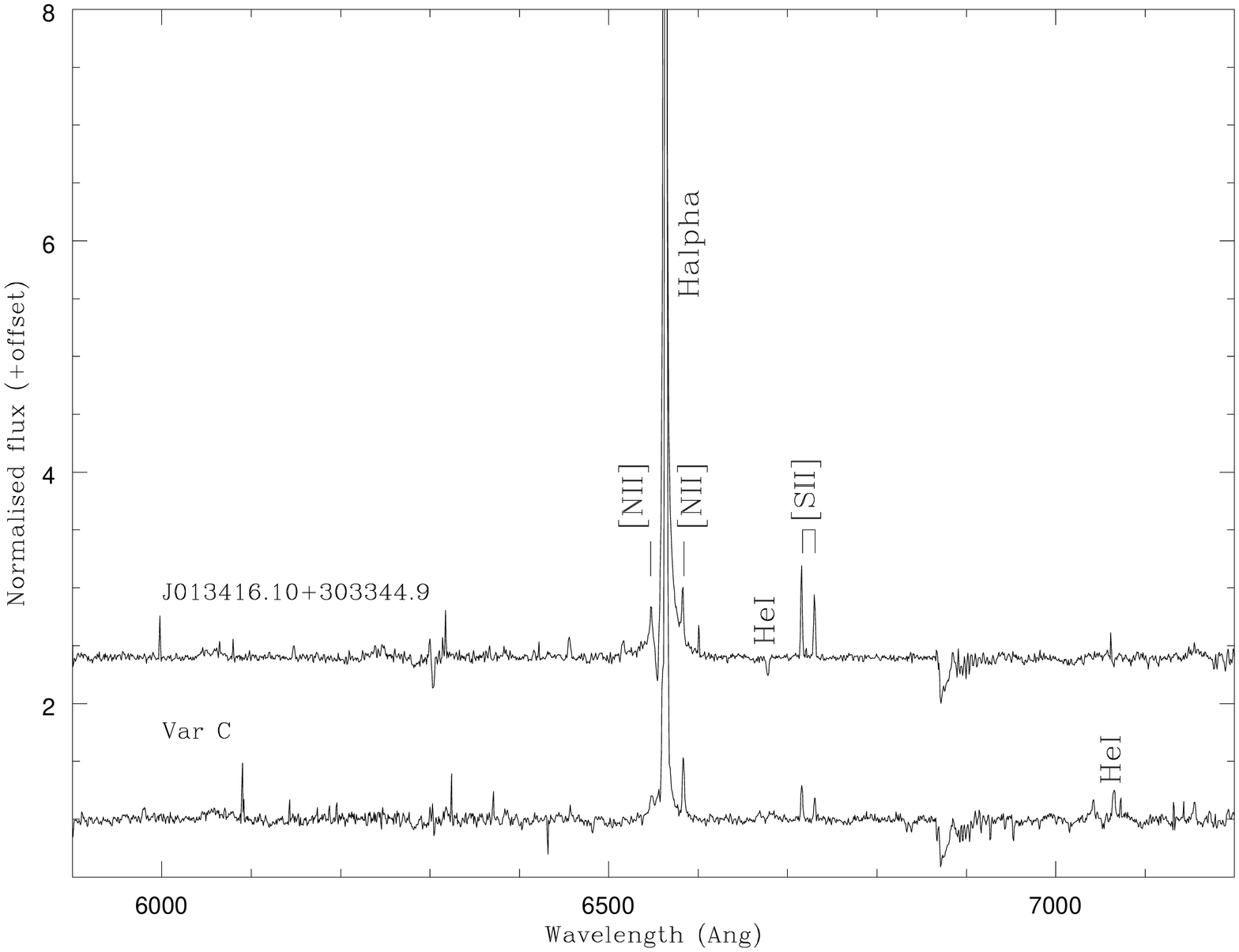}
\caption{Optical spectra of (candidate) LBVs. Note that the spectrum of M33 
Var B dates from 2003, and the spectrum of P Cygni has been  overplotted 
(red dashed line) for comparison, after having the 
resolution degraded to match.}
\end{figure*}


\begin{thebibliography}{}

\bibitem[2004]{abbott}
Abbott, J. B., Crowther, P. A., Drissen, L., et al. 2004, MNRAS, 350, 552
\bibitem[1989]{anders89} Anders, E. \&\ Grevesse, N. 1989, Geochimica et  Cosmochimica Acta, Vol. 53, 197
\bibitem[1999]{blocker}
Bl{\"o}cker, T., Balega, Y., Hofmann, K.-H., et al. 1999, A\&A, 348, 805
\bibitem[2007]{bolatto}
Bolatto, A. D., Simon, J. D., Stanimirov\'{i}c, S., et al. 2007, ApJ, 655, 212
\bibitem[2006]{bonano06} Bonanos, A.~Z., Stanek, K.~Z., Kudritzki, R.~P., et al. 2006, ApJ, 652, 313 
\bibitem[2009]{bonanos}
Bonanos, A. Z., Massa, D. L., Sewillo, M., et al. 2009, 
AJ, 138, 1003
\bibitem[2002]{bresolin}
Bresolin, F., Kudritzki, R.-P., Najarro, F., Gieren, W. \& 
Pietrzy\'{n}ski, G. 2002, ApJ, 577, L107
\bibitem[2011]{bresoli11} Bresolin, F.\ 2011, ApJ, 730, 129 
\bibitem[2008]{burggraf08}
Burggraf, B. \& Weis, K. 2008, ASPC, 388, 149
\bibitem[2011]{burggraf}
Burggraf, B., Weis, K., Bomans, D. J. \& Henze, M. 2011, BSRSL, 80, 440
\bibitem[2003]{clark03}
Clark, J. S., Egan, M. P., Crowther, P. A., et al. 2003, A\&A, 412, 185
\bibitem[2005a]{clark05a}
Clark, J. S., Larionov, V. M. \& Arkharov, A. 2005a, A\&A, 435, 239
\bibitem[2005b]{clark05b}
Clark, J. S., Neguerueula, I., Crowther, P. A. \& Goodwin, S. P. 2005b, A\&A, 
434, 949
\bibitem[2009]{clark09}
Clark, J. S., Crowther, P. A., Larionov, V. M., et al. 2009, A\&A,
507, 1555
\bibitem[2010]{clark10}
Clark, J.S., Ritchie, B. W. \& Negueruela, I. 2010, A\&A, 514, A87
\bibitem[2011a]{liege}
Clark, J. S., Arkharov, A., Larionov, V., et al.  2011a, BSRSL, 80, 361 
\bibitem[2011b]{clark11a}
Clark, J. S., Ritchie, B. W., Negueruela, I., et al. 2011b, A\&A, 531, A28
\bibitem[2012]{clark11b}
Clark, J. S., Najarro, F., Negueruela, I., et al. 2012, A\&A, in press [arXiv1202.3991]
\bibitem[2008]{cordiner}
Cordiner, M. A., Smith, K. T., Cox, N. L. J., et al. 2008, A\&A, 492, L5
\bibitem[1996]{corral}
Corral, L. J. 1996, AJ, 112, 1450
\bibitem[1997]{b517}
Crowther, P. A., Szeifert, Th., Stahl, O., \& Zickgraf, F.-J. 1997, A\&A, 318, 543
\bibitem[2009]{drout}
Drout, M. R., Massey, P., Meynet, G., Tokarz, S. \& Caldwell, N. 2009, ApJ, 
703, 441 
\bibitem[2002]{egan}
Egan, M. P., Clark, J. S., Mizuno, D. R., et al. 2002, ApJ, 572, 288
\bibitem[2005]{fabrika}
Fabrika, S., Sholukhova, O., Becker, T., et al. 2005, A\&A, 437, 217
\bibitem[1970]{fitzgerald}
Fitzgerald, M. P. 1970, A\&A, 4, 234
\bibitem[2011]{georgiev}
Georgiev, L. N., Koenigsberger, G. Hillier, D. J., et al. 2011, AJ, 142, 191
\bibitem[2009]{groh}
Groh, J. H., Hillier, D. J. \& Damineli, A. 2009, ApJ, 698, 1698
\bibitem[1995]{gummersbach}
Gummersbach, C. A., Zickgraf, F.-J. \& Wolf, B. 1995, A\&A, 302, 409
\bibitem[2010]{gvaramadze}
Gvaramadze, V. V., Kniazev, A. Y. \& Fabrika, S. 2010, MNRAS, 405, 1047
\bibitem[2007]{hadfield}
Hadfield, L. J., van Dyk, S. D., Morris, P. W., et al. 2007, MNRAS, 376, 248
\bibitem[2006]{hartman}
Hartman, J. D., Bersier, D. \& Stanek, K. Z. 2006, MNRAS, 371, 1405
\bibitem[1994]{herrero94}
Herrero, A., Lennon, D. J., Vilchez, J. M., Kudritzki, R. P. \& Humphreys, R. H. 1994, A\&A, 287, 885
\bibitem[1998]{h98}
Hillier, D. J. \& Miller, D. L. 1998, ApJ, 496, 407
\bibitem[1999]{h99}
Hillier, D. J. \& Miller, D. L. 1999, ApJ, 519, 354
\bibitem[1953]{hubble}
Hubble, E. \& Sandage, A. 1980, ApJS, 44, 319
\bibitem[1975]{humphreys75}
Humphreys, R. 1975, ApJ, 200, 426
\bibitem[1980]{humphreys80}
Humphreys, R. M. 1980, ApJ, 241, 587
\bibitem[1988]{humphreys88}
Humphreys, R. M., Leitherer, C., Stahl, O., Wolf, B. \& Zickgraf, F.-J.
1988, A\&A, 203, 306
\bibitem[1990]{humphreys90}
Humphreys, R. M., Massey, P. \& Freedman, W. L. 1990, AJ, 99, 84 
\bibitem[1994]{hd}
Humphreys, R. M. \& Davidson, K. 1994, PASP, 111, 1124
\bibitem[2006]{humphreys06}
Humphreys, R. M., Jones, T. J. \& Polomski, E. et al. 2006, AJ, 131, 2105
\bibitem[1996]{kaufer}
Kaufer, A., Stahl, O., Wolf, B., et al. 1996, A\&A, 305, 887
\bibitem[2011]{kehrig}
Kehrig, C., Oey, M. S. \& Crowther, P. A. 2011, A\&A, 526, A128
\bibitem[1985]{kenyon}
Kenyon, S. J. \& Gallagher, J. S. 1985, ApJ, 290, 542
\bibitem[2011]{khan}
Khan, R., Stanek, K. Z., Kochanek, C. S. \& Bonanos, A. Z. 2011, ApJ,
723, 43.
\bibitem[2010]{koen}
Koenigsberger, G., Gerogiev, L. \& Hillier, D. J. et al. 2010, AJ, 139, 2600
\bibitem[1995]{lamersdor}
Lamers, H. J. G. L. M. ASPC, 83, 176
\bibitem[1998]{lamerslbv}
Lamers, H. J. G. L. M., Bastiaanse, M. V., Aerts, C. \& Spoon, H. W. W. 1998, A\&A, 335, 605
\bibitem[1998]{lamers}
Lamers, H. J. G. L. M., Zickgraf, F.-J., de Winter, D., Houziaux, L. \& Zorec, J. 1998, A\&A, 340, 117 
\bibitem[2002]{lenorzer}
Lenorzer, A., de Koter, A. \& Waters, L. B. F. M. 2002, A\&A, 386, L5
\bibitem[2003]{lobel03}
Lobel, A., Dupree, A. K., Stefanik, R. P., et al. 2003, ApJ, 583, 923
\bibitem[2010]{maryeva}
Maryeva, O. \& Abolmasov, P. 2010, RMxAA, 46, 279
\bibitem[2011]{maryev11}
Maryeva, O. \& Abolmasov, P. 2012, MNRAS, 419, 1455
\bibitem[1995]{massey95}
Massey, P., Armandroff, T. E., Pyke, R., Patel, K. \& Wilson, C. D. 1995, AJ, 110, 2715
\bibitem[1996]{massey96}
Massey, P., Bianchi, L., Hutchings, J. B. \& Stecher, T. P. 1996,
ApJ, 469, 629
\bibitem[1998]{massey98}
Massey, P. \& Johnson, O. 1998, ApJ, 505, 793
\bibitem[2006]{massey06}
Massey , P., Olsen, K. A. G., Hodge, P. W., et al. 2006, AJ, 131, 2478
\bibitem[2007]{massey07}
Massey, P., McNeill, R. T. \& Olsen, K. A. G. 2007, AJ, 134, 2474
\bibitem[2007]{mcquinn}
McQuinn, K. B. W., Woodward, C. E., Willner, S. P., et al. 2007, ApJ, 664, 850
\bibitem[1996]{monteverde}
Monteverde, M. I., Herrero, A., Lennon, D. J. \& Kudritzki, R.-P.
1996, A\&A, 312, 24
\bibitem[2004]{morel}
Morel, T., Marchenko, S. V. \& Pati, A. K. 2004, MNRAS, 351, 552 
\bibitem[2000]{meynet}
Meynet, G. \& Maeder, A. 2000, A\&A, 361, 101 
\bibitem[1997]{najarro}
Najarro, F., Hillier, D. J. \& Stahl, O. 1997, A\&A, 326, 1117
\bibitem[2001]{paco01}
Najarro, F. 2001, ASP Conf. Ser. 233, P Cygni 2000: 400 Years of Progress, ed. M. de Groot \& 
C. Sterken (San Francisco, CA:ASP), 133
\bibitem[2011]{neugent}
Neugent, K. F. \& Massey, P. 2011, ApJ, 733, 123
\bibitem[1998]{oudmaijer}
Oudmaijer, R. D. 1998, A\&AS, 129, 541
\bibitem[2008]{patel}
Patel, M., Oudmaijer, R. D., Vink, J. S., et al. 2008, MNRAS, 385, 967
\bibitem[2005]{petrovic}
Petrovic, J., Langer, N. \& van der Hucht, K. A. 2005, A\&A, 435, 1013
\bibitem[2011]{polcaro}
Polcaro, V. F., Rossi, Viotti, R. F., et al. 2011, AJ, 141, 18
\bibitem[2008]{prieto}
Prieto, J. L. 2008, ATel., 1550, 1
\bibitem[2008]{prietoetal}
Prieto, J. L., Kistler, M. D. \& Thompson, T. A. 2008, ApJ, 681, L9
\bibitem[2009]{w243}
Ritchie, B. W., Clark, J. S., Neguereula, I. \& Najarro, F. 2009, A\&A, 507, 1597
\bibitem[1978]{romano}
Romano, G. 1978, A\&A, 67, 291
\bibitem[2011]{sana}
Sana, H. \&  Evans, C. J. 2011, IAUS, 272, 474
\bibitem[2009]{schuster}
Schuster, M. T., Marengo, M. \& Hora, J. L. 2009, ApJ, 699, 1423
\bibitem[2000]{shemmer}
Shemmer, O., Leibowitz, E. M. \& Szkody, P. 2000, MNRAS, 311, 698
\bibitem[2004]{sholukhova04}
Sholukhova, O., Fabrika, S., Roth, M. \& Becker, T. 2004, BaltA, 13, 156
\bibitem[2011]{sholukhova}
Sholukhova, O. N., Fabrika, S. N., Zharova, A. V., Valeev, A. F. \& Goranskij, 
V. P.  2011, AstBu, 66, 123 
\bibitem[2006]{shporer}
Shporer, A. \& Mazeh, T. 2006, MNRAS,370, 1429
\bibitem[2001]{smith01}
Smith, N., Humphreys, R. M., Davidson, K. et al., 2001, AJ, 121, 1111
\bibitem[2002]{smith02}
Smith, N., Gehrz, R. D., Hinz, P. M., et al. 2002, ApJ, 567, L77
\bibitem[2006]{smith06}
Smith, N. \& Owocki, S. P. ApJ, 645, L45
\bibitem[1992]{spiller}
Spiller, F. 1992, Ph. D. thesis, Heidelberg University 
\bibitem[2001]{stahl01}
Stahl, O., Jankovics, I., Kovacs, J., et al. 2001, A\&A, 375, 54 
\bibitem[2003]{stahl03}
Stahl, O., Gang, T., Sterken, C., et al. 2003, A\&A, 400, 279
\bibitem[2011]{stringfellow}
Stringfellow, G. S., Gvaramadze, V. V., Beletsky, Y. \& Kniazev, A. Y. 2011, to appear in the conference proceedings `Four Decades of Research on Massive Stars: A Scientific Meeting in 
Honour of Anthony F. J. Mofft', Laurent Drissen, Nicole St-Louis, Carmelle Robert and Anthony F. J. Moffat, eds. arXiv1112.2686
\bibitem[1996]{sromano}
Szeifert, T. 1996, in Wolf-Rayet stars in the framework of stellar evolution. ed. J. M. Vreux, A. Detal, D.
Fraipont-Caro, E. Gosset and G. Rauw (Liege:Univerite de Liege, Institut d'Astrophysique), p. 459
\bibitem[1996]{szeifert}
Szeifert, Th., Humphreys, R. M., Davidson, K., et al. 1996, A\&A, 314, 131
\bibitem[2003]{szeifert03}
Szeifert, Th., Kaufer, A., Crowther, P. A. \& Stahl, O. 2003, in A Massive Star Odyssey: From Main Sequence to Supernova, Proceedings
of IAU Symposium 212, ed. K. van der Hucht, A. Herrero and C. Esteban, (San Francisco, CA:ASP), p. 243
\bibitem[2009]{thompson}
Thompson, T. A., Prieto, J. L., Stanek, K. Z., et al. 2009, ApJ, 705, 1364 
\bibitem[2005]{urbaneja05}
Urbaneja, M. A., Herrero, A., Kudritzki, R.-P., et al. 2005, ApJ, 635, 311
\bibitem[2011]{urbaneja11}
Urbaneja, M. A., Herrero, A., Lennon, D. J., Corral, L. J. \&  Meynet, G. 2011, ApJ, 735, 39
\bibitem[2009]{valeev09}
Valeev, A. F., Sholukhova, O. \& Fabrika, S. 2009, MNRAS, 396, L21
\bibitem[2010]{valeev10}
Valeev, A. F., Sholukhova, O. N. \& Fabrika, S. N. 2010, AstBu, 65, 381
\bibitem[1975]{83}
van den Berg, S., Herbst, E. \& Kowal, C. T. 1975, ApJS, 29, 303
\bibitem[2001]{vanD}
van Dokkum, P. G. 2001, PASP, 113, 1420
\bibitem[2001]{vG}
van Genderen, A. M. 2001, A\&A, 366, 508
\bibitem[1998]{alpha}
van Leeuwen, F., van Genderen, A. M. \& Zegelaar, I. 1998, A\&AS, 128, 117
\bibitem[2006]{viotti}
Viotti, R. F., Rossi, C., Polcaro, V. F., et al. 2006, A\&A, 458, 225
\bibitem[2000]{voors}
Voors, R. H. M., Waters, L. B. F. M., de Koter, A., et al. 2000, A\&A, 356, 
501 
\bibitem[2010]{wachter}
Wachter, S., Mauerhan, J. C., van Dyk, S. D., et al. 2010, AJ, 139, 2330
\bibitem[1990]{walborn90}
Walborn, N. R. \& Fitzpatrick, E. L. 1990, PASP, 102, 379
\bibitem[2000]{walborn00}
Walborn, N. R. \& Fitzpatrick, E. L. 2000, PASP, 122, 50
\bibitem[2008]{walborn08}
Walborn, N. R., Stahl, O. \& Gamen, R. G. 2008, APJ, 683, L33 
\bibitem[1994]{wegner}
Wegner, W. 1994, MNRAS, 270, 229
\bibitem[1981]{wolf81}
Wolf, B., Appenzeller, I., Stahl, O. 1981, A\&A, 103, 94
\bibitem[1988]{wolf}
Wolf, B., Stahl, O., Smolinski, J. \& Casatella, A. 1988, A\&AS, 74, 239
\bibitem[1989]{wolf89}
Wolf, B. 1989, A\&A, 217, 87
\bibitem[1985]{zickgraf85}
Zickgraf, F.-J., Wolf, B., Stahl, O., Leitherer, C. \& Klare, G. 1985, A\&A, 143, 421
\bibitem[1986]{zickgraf86}
Zickgraf, F.-J., Wolf, B., Stahl, O., Leitherer, C. \& Appenzeller, I. 1986, A\&A, 163, 119
\bibitem[2006]{zickgraf06}
Zickgraf, F.-J. 2006, in Stars with the B[e] phenomenon. ASP Conference Series Vol. 355 (San Francisco, CA:ASP) p. 
211

\end{thebibliography}
\end{document}